\documentclass[%
 superscriptaddress,
 twocolumn,
 amsmath,amssymb,
 aps,
 prl,
]{revtex4-2}

\usepackage{graphicx}
\usepackage{dcolumn}
\usepackage{bm}
\usepackage{natbib}

\usepackage{marginnote}
\usepackage{bm}

\usepackage{float}
\usepackage{xcolor}
\usepackage{soul}

\usepackage{hyperref}
\hypersetup{
  colorlinks   = true, 
  urlcolor     = blue, 
  linkcolor    = blue, 
  citecolor   = blue 
}

\makeatletter
\def\maketitle{
\@author@finish
\title@column\titleblock@produce
\suppressfloats[t]}
\makeatother

\newcommand{\target}{sensing}
\newcommand{\control}{detected}

\begin{document}

\preprint{APS/123-QED}

\title{Coulomb sensing of single ballistic electrons}

\author{J. D. Fletcher} \thanks{These authors contributed equally to this work.}
\affiliation{National Physical Laboratory, Hampton Road, Teddington TW11 0LW, United Kingdom}
\author{W. Park} \thanks{These authors contributed equally to this work.} 
\affiliation{Department of Physics, Korea Advanced Institute of Science and Technology, Daejeon 34141, Korea}
\affiliation{Korea Research Institute of Standards and Science, Daejeon 34113, Korea}
\author{P. See} 
\affiliation{National Physical Laboratory, Hampton Road, Teddington TW11 0LW, United Kingdom}
\author{J. P. Griffiths} 
\affiliation{Cavendish Laboratory, University of Cambridge, J. J. Thomson Avenue, Cambridge CB3 0HE, United Kingdom}
\author{G.~A.~C.~Jones} 
\affiliation{Cavendish Laboratory, University of Cambridge, J. J. Thomson Avenue, Cambridge CB3 0HE, United Kingdom}
\author{I. Farrer}\thanks{Present address: School of Electrical and Electronic Engineering, University of Sheffield, Sheffield, UK} 
\affiliation{Cavendish Laboratory, University of Cambridge, J. J. Thomson Avenue, Cambridge CB3 0HE, United Kingdom}
\author{D. A. Ritchie} 
\affiliation{Cavendish Laboratory, University of Cambridge, J. J. Thomson Avenue, Cambridge CB3 0HE, United Kingdom}
\author{H.-S. Sim} \email[]{hs\_sim@kaist.ac.kr}
\affiliation{Department of Physics, Korea Advanced Institute of Science and Technology, Daejeon 34141, Korea}
\author{M. Kataoka} \email[]{masaya.kataoka@npl.co.uk}
\affiliation{National Physical Laboratory, Hampton Road, Teddington TW11 0LW, United Kingdom}

\date{\today}

\begin{abstract}
While ballistic electrons are a key tool for applications in sensing and flying qubits, sub-nanosecond propagation times and complicated interactions make control of ballistic single electrons challenging. Recent experiments have revealed Coulomb collisions of counterpropagating electrons in a beam splitter, giving time resolved control of interactions between single electrons. Here we use remote Coulomb interactions to demonstrate a scheme for sensing single ballistic electrons. We show that interactions are highly controllable via electron energy and emission timing. We use a weakly-coupled `sensing' regime to characterise the nanoscale potential landscape of the beam splitter and the strength of the Coulomb interaction, and show multi-electron sensing with picosecond resolution.
\end{abstract}

\maketitle

{\it Introduction.--} Control and detection of ballistic electrons is key to electron quantum optics~\cite{FeveScience2007}, quantum electrical metrology \cite{giblin2012towards}, flying qubit technology~\cite{edlbauer2022semiconductor,thiney2022flight}, and signal sensing~\cite{johnson2017ultrafast}. However, charge-sensing schemes using confined electrons~\cite{vandersypen2004real,Field_PhysRevLett.70.1311,schoelkopf1998radio} have insufficient time resolution to detect propagating electrons. In quantum Hall edge channels~\cite{Kataoka2016,fletcher2019continuous} or narrow wires~\cite{roussely2018unveiling} electrons propagate in times (0.01-1~ns) shorter than a typical charge sensor readout time $10^{-6}$~s~\cite{noiri2020radio,reilly2007fast}. Although there are various approaches, including AC current detection~\cite{FeveScience2007}, synchronous partitioning~\cite{Fletcher2013}, shot noise~\cite{ubbelohde2015partitioning}, charge capture~\cite{Freise2020}, qubit sensors~\cite{thiney2022flight} and ``which path'' detectors~\cite{buks1998dephasing,sprinzak2000controlled}, a more direct effect with high time resolution would be preferable.

An idealised sensing scheme is shown in Fig.~\ref{fig:figure1}. The Coulomb repulsion from a `\control{}' electron changes a `\target{}' electron through a nearby barrier. This is akin to a conventional charge sensor~\cite{Field_PhysRevLett.70.1311} where the object being detected is a single moving charge but the detector current is also a single ballistic electron. Crucially, the delay between injection times $t_1$ and $t_2$ enables a time-selective interaction. This is analogous to techniques used in ultrafast optics in which one beam is analyzed using another beam delayed in time~\cite{krokel1989subpicosecond, weiner2011ultrafast, hunter2015chip}. Recently, weakly-screened Coulomb repulsion between single electrons colliding at a beam splitter has been reported~\cite{ryu2022partition, Fletcher2023, ubbelohde2023two, wang2023coulomb}. These interactions provide a mechanism for two-qubit gates in flying qubits~\cite{edlbauer2022semiconductor} but they also suggest that time-selective sensing schemes like that in Fig.~\ref{fig:figure1} may be achievable.

\begin{figure}[b]
	\includegraphics[width=0.3\textwidth]{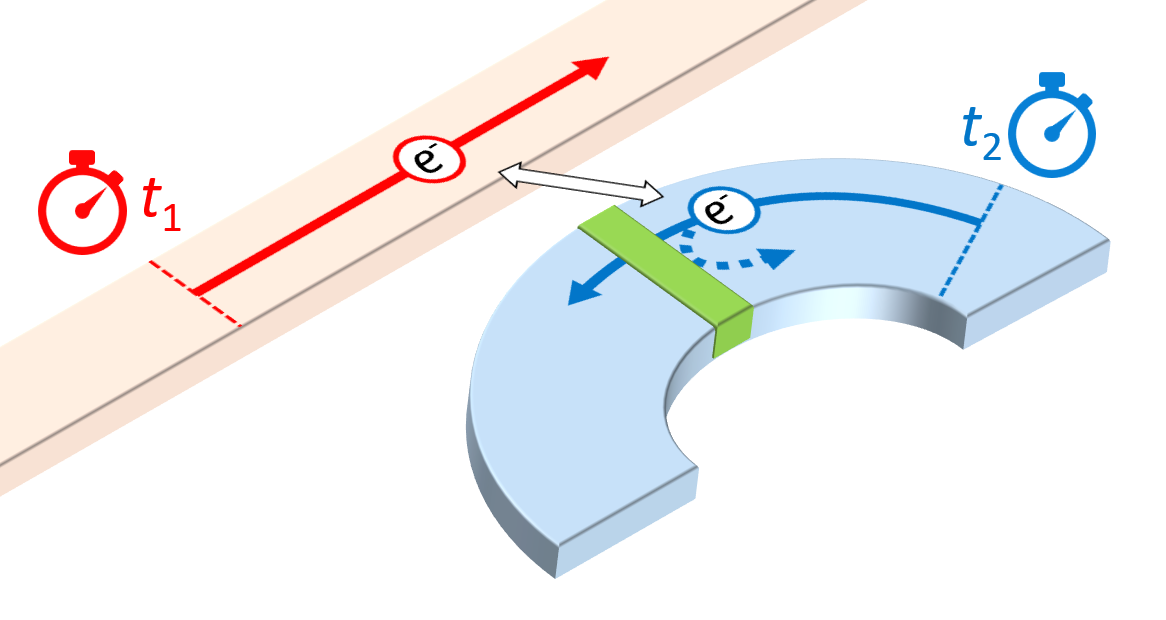}
	\caption{Time-resolved sensing of ballistic single electrons. The transmission of a `\target{}' electron (blue arrows) through a barrier (green) in the lower channel is perturbed by Coulomb interaction with a `\control{}' electron in the upper channel (red arrow).}
	\label{fig:figure1}
\end{figure}

In this work, we control and measure a long-range Coulomb interaction between two ballistic electrons from on-demand sources propagating in quantum Hall channels. This enables a system for sensing single ballistic electrons on picosecond time scales. Using a microscopic model we find that timing resolution is determined both by the potential landscape in which the interaction occurs and by the emission distribution of the electron pumps. Analysis of collisions at different relative energies and injection times separates these effects and enables extraction the geometrical parameters of the potential barrier. As an example application we use this scheme to study multi-electron bunches emitted from a single pump~\cite{Fletcher2013}. We also discuss the ultimate limitations of single electron sensing in this mode.

\begin{figure}[t]
\includegraphics[width=0.45\textwidth]{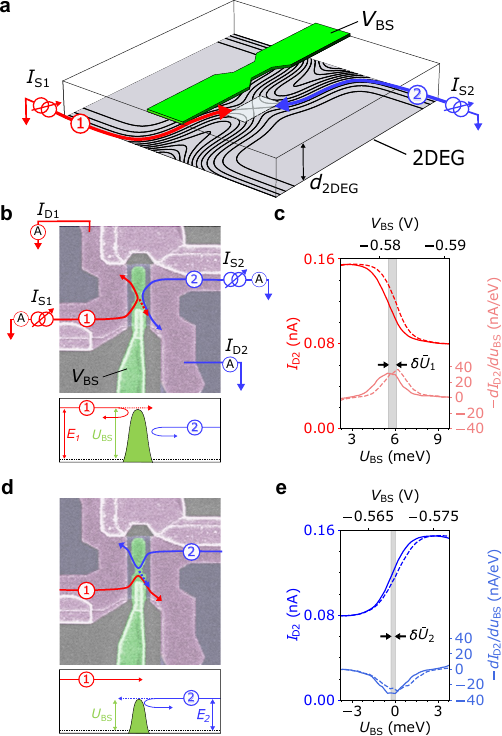}
\caption{%
Charge sensing of ballistic electrons.
    (a) Single electrons from sources S1 and S2 travel in edge states meeting at a beam splitter controlled by gate voltage $V_{\rm BS}$.
    (b) Electron trajectories sketched on image of device geometry (upper) and energy level sketch (lower). Electrons $i=1,2$ with energy $\bar{E}_i$ meet at beam splitter with height $U_{\rm BS}$ (green). For $U_{\rm BS} \simeq \bar{E}_1 > \bar{E}_2$, electron $i=1$ is partitioned (reflected or transmitted) while $i=2$ is always reflected.
    (c) Detector current $I_{\rm D2} (U_{\rm BS})$ (upper) and transconductance $dI_{\rm D2}/dU_{\rm BS}$ (lower) when $U_{\rm BS} \simeq \bar{E}_1$ and $\Delta \bar{E} \equiv \bar{E}_2 - \bar{E}_1 = -6.0$ meV [depicted in (b)]. $\Delta{\bar{t}} \equiv \bar{t}_2 - \bar{t}_1 = 0$~ps (solid line) and $\Delta{\bar{t}} = -20$~ps (dashed line).
    (d), (e) as (b), (c) but for $U_{\rm BS} \simeq \bar{E}_2 < \bar{E}_1$, $i=2$ is partitioned (reflected or transmitted) while $i=1$ is always transmitted.
}
\label{fig:figure2}
\end{figure}

{\it Experimental setup.--} We use independent sources S1 and S2 of high energy electrons~\cite{Fletcher2013,Kataoka2016,giblin2012towards}. Each source S$i$ ($i=1,2$) emits a single electron $i$ at frequency $f = 500$~MHz, giving source currents $I_{\rm S1}, I_{\rm S2} = ef$ [Fig.~\ref{fig:figure2}(a,b)].
Electron $i$ has a mean injection energy $\bar{E}_i$ and a mean injection time $\bar{t}_i$ (overbar denotes the mean value). 
The source parameters~\cite{supp}, synchronisation techniques, and parameters of the incident wave packets are discussed in Refs.~\cite{Fletcher2023}. The injected electrons $i=1,2$ travel along quantum Hall edges [Fig.~\ref{fig:figure2}(a)] of a GaAs two-dimensional electron gas (2DEG) in a perpendicular magnetic field $B= 10$~T. They approach each other at a beam splitter with potential barrier height $U_{\rm BS}$ controlled by a gate voltage $V_{\rm BS}$; Fig.~\ref{fig:figure2}(b) shows the case when electron 1 is at partial transmission and Fig.~\ref{fig:figure2}(d) when electron 2 is at partial transmission. The scaling $V_{\rm BS} \simeq 2.2 U_{\rm BS}$ is obtained as in Ref.~\cite{taubert2011relaxation}.

\begin{figure}[t]
	\includegraphics[width=0.45\textwidth]{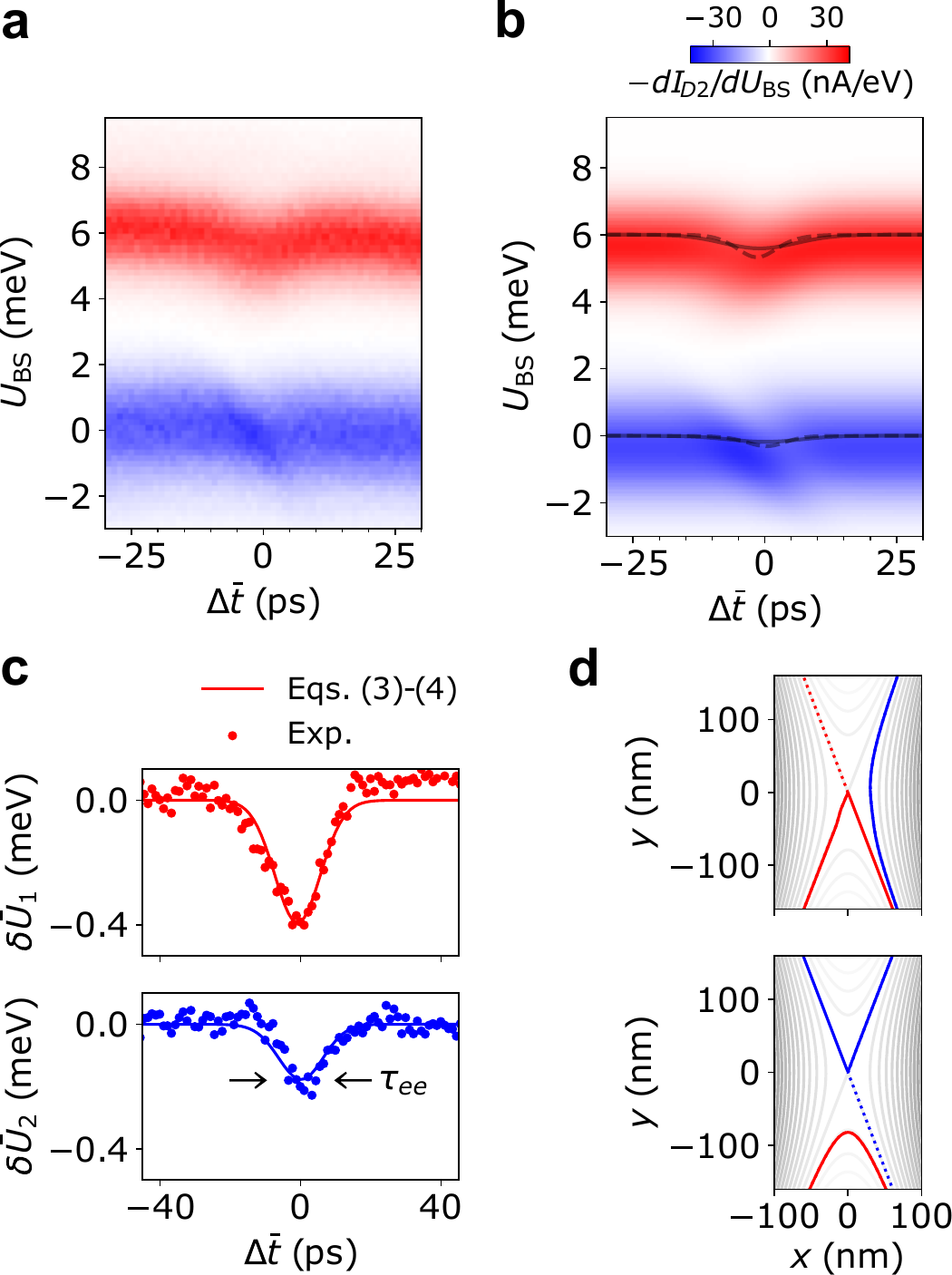}
	\caption{%
		Readout of the potential landscape of the beam splitter and the Coulomb interaction strength in the sensing regime of $\Delta \bar{E} = - 6$ meV. (a) Experimental and (b) numerically computed $dI_{\rm D2}/dU_{\rm BS}(U_{\rm BS},\Delta \bar{t})$. Solid and dashed lines in (b) denote $\delta \bar{U}_i$ [Eq.~\eqref{eq:deltabarE_th}] and $\delta U_i$ [Eq.~\eqref{eq:deltaE_th}], respectively. 
        (c) Experimental $\delta \bar{U}_i$ (symbols) obtained from the data $dI_{\rm D2}/dU_{\rm BS}$ in (a) using Eq.~\eqref{eq:barE}. It is compared to $\delta \bar{U}_i$ (line) obtained from Eqs.~\eqref{eq:deltabarE_th} and \eqref{eq:deltaE_th}. (d) Classical trajectories of electrons $i=1$ (red) and $i=2$ (blue) for $\Delta \bar{t} = 0$ at $U_{\rm BS} \simeq \bar{E}_1$ (upper panel) and $U_{\rm BS} \simeq \bar{E}_2$ (lower panel). Contours depict beam-splitter equipotential lines. In all computations, $\omega_x = 5.8$ ps$^{-1}$, $\omega_y = 2.2$ ps$^{-1}$, and $\beta=95$ meV$\cdot$nm [taken as fitting parameters for comparison in (c)].}
	\label{fig:figure3}
\end{figure}

{\it Coulomb sensing.--} The electrons initially follow the equipotential lines of their injection energy $\bar{E}_i$ as in Fig.~\ref{fig:figure2}(a). Transmission through the barrier depends on this energy compared to that of the saddle point, $U_{\rm BS}$~\cite{Fletcher2013,waldie2015measurement,ubbelohde2015partitioning,fletcher2019continuous}. The detector current $I_{\rm D2}$ for a given $U_{\rm BS}$ and the relative injection time $\Delta \bar{t} \equiv \bar{t}_2 - \bar{t}_1$ is obtained by averaging over $20$ ms ($10^7$ cycles at a repetition rate of $500$~MHz). When the beam-splitter height $U_{\rm BS}$ passes through a threshold value of $\bar{U}_i$, the transmission probability of electron $i$ through the beam splitter decreases from 1 to 0. The detector current transconductance $\frac{dI_\textrm{D2}}{dU_{\rm BS}} (U_{\rm BS})$ shows a peak/dip at $U_{\rm BS} \simeq \bar{U}_i$. Figure~\ref{fig:figure2}(c) shows $U_{\rm BS} \simeq \bar{U}_1$ and Figure~\ref{fig:figure2}(e) where $U_{\rm BS} \simeq \bar{U}_2$ (peaks/dips are inverted in $\frac{dI_\textrm{D1}}{dU_{\rm BS}}$). The position of $\bar{U}_i$ is modified by $\delta\bar{U}_i$ due to an additional Coulomb repulsion when electrons arrive at the barrier simultaneously (solid lines, $\bar{t}_1 \simeq \bar{t}_2$) compared to the reference measurement where electrons arrive separately~\cite{Fletcher2023} (dashed lines, $\bar{t}_1 = \bar{t}_2 + 20$ ps). Changes in the transmitted current of the \target{} electron indicate the presence or absence of a \control{} electron in the time window sampled by the interaction. This gives the time-selectivity of the scheme in Figure~\ref{fig:figure1}. We measure the sensitivity and time resolution of our scheme then relate this to the microscopic picture of the scattering, including the shape of the barrier potential and the effect of the electron pump emission distribution~\cite{fletcher2019continuous}.

The shift in the threshold barrier height $\delta \bar{U}_i$ depends on the relative injection time $\Delta{\bar{t}} \equiv \bar{t}_2 - \bar{t}_1$ with a peak value near $\Delta \bar{t} = 0$ (synchronised arrival). This can be seen in the raw transconductance color map $dI_{\rm D2}/dU_{\rm BS}$ in Fig.~\ref{fig:figure3}(a). In this map, the upper feature (red) occurs when the \target{} electron is $i=1$, the \control{} electron is $i=2$, and the beam-splitter height varies around $U_{\rm BS} \simeq \bar{E}_1$ [Fig.~\ref{fig:figure2}(b)]. For the lower (blue) feature \target{} and \control{} electrons are swapped ($i=2$ and $i=1$ respectively) when $U_{\rm BS} \simeq \bar{E}_2$ [Fig.~\ref{fig:figure2}(d)].

To quantify the sensitivity of the Coulomb interaction we measure $\bar{U}_i$ at each time delay using the weighted average
\begin{equation}
\bar{U}_i(\Delta \Bar{t}) \equiv \frac{\int_i dU_{\rm BS} \, U_{\rm BS} \, dI_{\rm D2} /dU_{\rm BS}}{\int_i dU_{\rm BS} \, dI_{\rm D2}/dU_{\rm BS}}.
	\label{eq:barE}
\end{equation}
and Coulomb-driven change is quantified by
 \begin{equation}
 	\delta \bar{U}_i(\Delta \Bar{t}) \equiv \bar{U}_i(\Delta \Bar{t}) - \bar{U}_i(\Delta \Bar{t} \to \infty)
 	\label{eq:deltabarE}
 \end{equation}
for given $\Delta \bar{E}$. 
Extracted values of $\delta \bar{U}_i$, corrected for a small cross talk effect between the pump driving gates and the beam splitter~\cite{supp}, are shown in Fig.~\ref{fig:figure3}(b). 

The Coulomb-driven shift $\delta \bar{U}_i$ reaches a peak value  $\max |\delta \bar{U}_i| \simeq 0.2$ and $0.4$~meV (the value varies for $i$), a small shift (less than 1\%) on the scale of the injection energy~\cite{Fletcher2013}. The Coulomb repulsion effect appears in a region of $\Delta \bar{t}$ spanning only $\tau_{ee} \simeq 6~$ps illustrating the extremely time-selective nature of this technique and the high effective bandwidth.

We use adjustment of the injection energy\cite{Fletcher2013} to extract the information about the electron trajectory from the strength of Coulomb repulsion. In Fig.~\ref{fig:figure3}(a) the energy separation $\Delta \bar{E} \equiv \bar{E}_2 - \bar{E}_1 = - 6$~meV. Reducing $\Delta \bar{E}$ increases the visibility of the Coulomb interaction feature (peak value of $\delta \bar{U}_i$) as  the electron trajectories approach each other more closely. Combined with a microscopic model, this data provides key information about the Coulomb interaction and explains the sensitivity and time resolution of this ballistic single-electron sensing technique.

\begin{figure}[t]
	\includegraphics[width=0.45\textwidth]{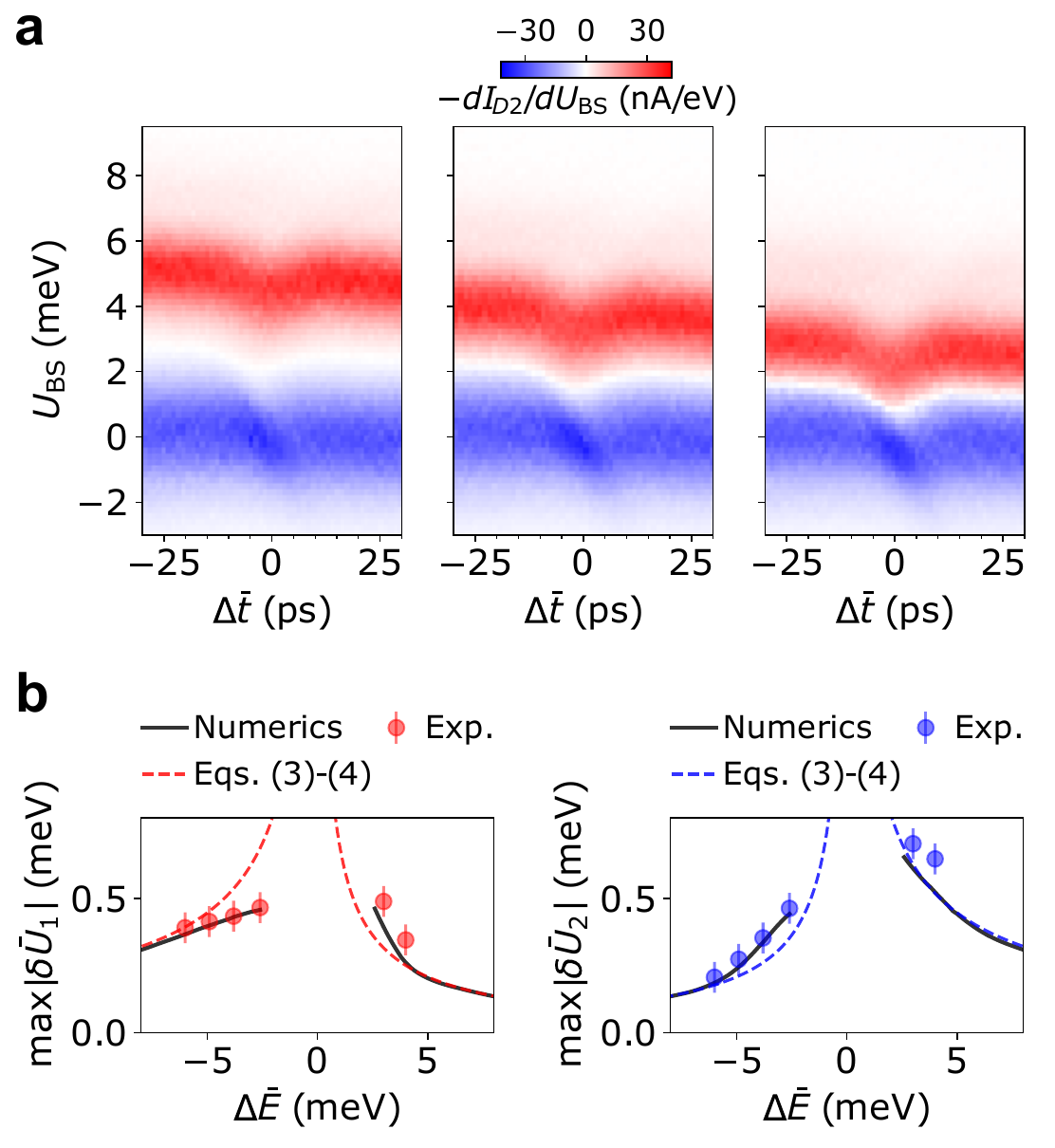}
	\caption{%
         Coulomb sensing for different injection energy differences $\Delta \bar{E}$. (a) Transconductance maps $dI_{\rm D2}/dU_{\rm BS}(U_{\rm BS},\Delta \bar{t})$ for $\Delta \bar{E} = \bar{E}_2 - \bar{E}_1 = -4.9, -3.8, -2.6$ meV (from left to right). (b) $\max{|\delta \bar{U}_i|}$ as a function of $\Delta \bar{E}$. Circles are obtained from experimental data. The error bars indicate two-sigma regions, where sigma represents the fluctuation of $\delta \bar{U}_i$ in the experiment~\cite{supp}. Solid lines are obtained by numerical integration of computed $dI_{\rm D2}/dU_{\rm BS}$, and dashed curves are from Eqs.~\eqref{eq:deltabarE_th} and \eqref{eq:deltaE_th}. In computations, $\omega_x = 5.8$ ps$^{-1}$, $\omega_y = 2.2$ ps$^{-1}$, and $\beta=95$ meV$\cdot$nm.} 
	\label{fig:figure4}
\end{figure}

{\it Semiclassical model.--} We compare our data to a model using $E \times B$ electron classical trajectories~\cite{Fletcher2023, park2023measuring, pavlovska2023collision} combined with the effect of the Wigner distribution of the incident electrons (see Ref.~\cite{supp}). We show below that this gives quantitative agreement with our experimental data after choosing parameters that capture the potential landscape of the beam splitter and the effective strength of the Coulomb interaction (other parameters are taken from experiments). 

The beam splitter is described by a saddle potential~\cite{Fertig-1987-1}, $\mathcal{U}_{2D}(x,y) = U_{\rm BS} - m^* \omega_x^2 x^2/2 + m^* \omega_y^2 y^2/2$, with curvatures $\omega_x$ and $\omega_y$, and where $m^*$ is the electron effective mass. The mutual Coulomb repulsion $\mathcal{U}_{ee}$ is dependent on the distance $r$ between electrons but also on charge screening~\cite{barthel2010fast}. In our device the beam splitter 2DEG region is depleted but there are metallic surface gates at $d_{\rm 2DEG} \simeq 90$~nm above the 2DEG. The Coulomb interaction $\mathcal{U}_{ee}(r)$ between electrons $i=1,2$ is screened ($\mathcal{U}_{ee} \rightarrow 0$) by these surface gate when $r > d_{\rm 2DEG}$ [Fig.~\ref{fig:figure2}(a)] but is unscreened ($\mathcal{U}_{ee}\propto 1/r$), when they approach each other at the beam splitter for $r \lesssim d_{\rm 2DEG}$. We take $\mathcal{U}_{ee}(r) = \beta[ 1/r - 1/\sqrt{r^2 + 4d_{\rm 2DEG}^2}]$ which uses the method of image charges~\cite{SkinnerPhysRevB.82.155111} to account for the screening by the gate with all other effects. The effective dielectric constant and any other screening effects are accounted for by an interaction strength $\beta$. 

{\it Beam splitter parameters.--}  We describe our method to estimate the parameters of the beam splitter potential $\mathcal{U}_{2D}$ and the Coulomb strength $\beta$ using $\delta U_i$ and the corresponding analytical equations [Eqs.~\eqref{eq:deltabarE_th} and \eqref{eq:deltaE_th}] introduced later. A calculation with optimised parameters, $\omega_x, \omega_y$, and $\beta$ shows excellent agreement with the data [Fig.~\ref{fig:figure3}(c)]. These equations are applied within the regime where the classical model can be computed in the perturbative limit, where $|\Delta \bar{E}|$ is much larger than the Coulomb effect ($\delta U_i \leq 1$ meV) and energy broadening ($\sigma_{Ei} \simeq 1$~meV). In this regime we derive~\cite{supp}
\begin{equation}
	\delta \bar{U}_{i}(\Delta \Bar{t}) \simeq \int^{\infty}_{-\infty} dt_1 \int^{\infty}_{-\infty} dt_2 \, \rho_1(t_1) \rho_2(t_2) \delta U_i(\Delta t),
	\label{eq:deltabarE_th}
\end{equation}
where $\Delta t \equiv t_2 - t_1$ and $\rho_i$ is the injection time distribution of electron $i$. $\delta U_i$ is the difference in the classical potential height threshold of the beam splitter (to block electron $i$) created by the Coulomb interaction $\mathcal{U}_{ee}$. Using the saddle potential $\mathcal{U}_{2D}$ and treating $\mathcal{U}_{ee}$ perturbatively, we derive~\cite{supp}
\begin{equation}
	\delta U_i(\Delta t) \simeq -\dfrac{\beta}{\sqrt{|\Delta \bar{E}|}} f_i(\Delta t;\omega) g_i(\omega_x,\omega_y) + \text{screening term}.
	\label{eq:deltaE_th}
\end{equation}
$\omega = \omega_x \omega_y / \omega_c$ is the characteristic frequency of the beam splitter saddle potential, $\omega_c$ is the cyclotron frequency, $f_1(\Delta t;\omega) \equiv e^{- \omega \Delta t}/(1+e^{-\omega \Delta t})^{3/2}$, $f_2(\Delta t;\omega) \equiv e^{\omega \Delta t}/(1+e^{\omega \Delta t})^{3/2}$, and $g_i$ is the geometric factor whose ratio $g_1/ g_2$ depends only on the anisotropy factor $\omega_x/\omega_y$ of the saddle potential. The first and second terms of Eq.~\eqref{eq:deltaE_th} are from the unscreened ($1/r$) and screened parts of $\mathcal{U}_{ee}$. Equations~\eqref{eq:deltabarE_th} and \eqref{eq:deltaE_th} agree with the numerical calculations in the case of the value of $\Delta \bar{E} = -6$ meV [Fig.~\ref{fig:figure3}], the regime used to estimate parameters of the beam splitter as described below. The interacting trajectories in this case [Fig.~\ref{fig:figure3}(d)] do not deviate markedly from the equipotential lines indicating the validity of equation~\eqref{eq:deltaE_th}.

We analyze how the saddle potential parameters and the Coulomb strength $\beta$ determine $\delta \bar{U}_i$ using Eqs.~\eqref{eq:deltabarE_th}--\eqref{eq:deltaE_th}. The injection time distributions~\cite{Fletcher2023} smear out the measured response increasing the values of $\tau_{ee}$. From Eq.~\eqref{eq:deltabarE_th} we find that $\tau_{ee} = \sqrt{\sigma_{t1}^2 + \sigma_{t2}^2 + \tau_{\omega}^2}$, a mixture of the temporal wavepacket width $\sigma_{ti}$ of electron $i$ and the characteristic time $\tau_{\omega} \simeq 1/\omega = \omega_c / (\omega_x \omega_y)$ of the saddle potential. Using the measured values of $\tau_{ee} \simeq 6$ ps, $\sigma_{t1} = 1.7$ ps, and $\sigma_{t2} = 5.2$ ps (and $\omega_c \simeq 26.25$ ps$^{-1}$ at $B = 10$~T) we estimate $\tau_\omega \sim 2$~ps and $\omega_x \omega_y \simeq 11 \text{ ps}^{-2}$. For accurate time-resolved sensing, the temporal width of the \target{} electron and $\tau_{\omega}$ should be minimized.

Equation~\eqref{eq:deltaE_th} indicates that anisotropy in the saddle potential $\omega_x / \omega_y$ can be deduced from $\max{\delta U_1} / \max{\delta U_2}$. This geometrical effect can be understood from the different proximity of the \control{} electron trajectory to the saddle point centre where the \target{} electron transmission is determined. For $\omega_x > \omega_y$ the turning point of $i=1$ [upper panel of Fig.~\ref{fig:figure3}(d)] is closer to the saddle point than that of $i=2$ [lower panel], resulting in $\max{ \delta U_1}/\max{\delta U_2} > 1$. By symmetry, these effects are reversed for $\Delta \bar{E}>0$. Experimentally $\max{\delta \bar{U}_{1}}/\max{\delta \bar{U}_{2}} \simeq 1.8$ indicates $\omega_x > \omega_y$ in our device. This is shown in the sketch of the equipotential lines of the beam splitter in Fig.~\ref{fig:figure2}(a). The interaction strength $\beta$ is estimated from $\max{\delta \bar{U}_i}$ for given $\Delta \bar{E}$. The first term of Eq.~\eqref{eq:deltaE_th} is proportional to $1/ \sqrt{|\Delta \bar{E}|}$, which is a signature of the unscreened long-range $1/r$ interaction. The proximity of electron trajectories $i = 1, 2$ changes as the relative injection energy $|\Delta \bar{E}|$ is tuned.

An estimation~\cite{supp} based on a least square deviation between the experimental data of $\delta \bar{U}_i$ and Eqs.~\eqref{eq:deltabarE_th} and \eqref{eq:deltaE_th} (including also its second term) provides $\omega_x = 5.8 \pm 0.6$ ps$^{-1}$, $\omega_y = 2.2 \pm 0.2$ ps$^{-1}$, and $\beta = 95 \pm 4 $ meV$\cdot$nm [Fig.~\ref{fig:figure3}]. This value of $\beta$ is close to $\beta \simeq 112$ meV$\cdot$nm for unscreened interactions in GaAs. While the pump emission distribution reduces the maximum detected $\delta \bar{U}_i$ by a factor $\simeq 2$ here [see lines in Fig.~\ref{fig:figure3}(b)], this is accounted for in the model.

{\it Injection energy dependence.--} We check the robustness of these parameters and the Coulomb potential $\mathcal{U}_{ee}(r)$ by varying the mean energy difference $\Delta \bar{E}$ (hence varying the inter-particle distance $r$). Experimental data at $\Delta \bar{E} = -4.9, -3.8, -2.6$ meV are shown in Fig.~\ref{fig:figure4}(a) and for $\Delta \bar{E} = 3, 4$ meV (see Supplementary Material~\cite{supp}). Although the behaviour at smaller $\Delta \bar{E}$ no longer necessarily coincides with the simple  $1/\sqrt{\Delta \bar{E}}$ form [dashed lines in Fig.~\ref{fig:figure4}(b)] the values of $\max{\delta \bar{U}_i}$ (squares) agree with numerical computations (solid lines) (see Ref.~\cite{supp} for detailed comparisons). The asymmetric behavior of $\delta \bar{U}_i$ with respect to the sign of $\Delta \bar{E}$ (which is mirrored for $i=1,2$) arises from the anisotropic potential with $\omega_x > \omega_y$~\cite{supp}. Understanding these geometrical effects is essential for the integration and optimisation of ballistic electron detection schemes.

\begin{figure}[!t]
\includegraphics[width=0.45\textwidth]{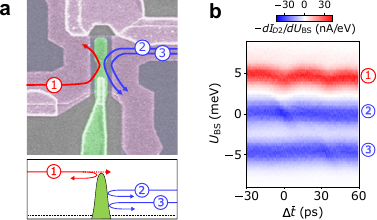}
\caption{Selective sensing among multiple electrons.
(a) Injection of a single electron from source S1 and two electrons (2 and 3) from source S2. 
(b) Transconductance $dI_{\rm D2}/dU_{\rm{BS}}$ maps show Coulomb sensing between electrons 1 and 2 at $\Delta \bar{t} =0$ and between electrons 1 and 3 at $\Delta \bar{t} \simeq 35$~ps. 
}
\label{fig:figure5}
\end{figure}

{\it Sensing sequential electrons.--} One use of this scheme is to perform time-resolved sensing of a multi-electron cluster. In Fig.~\ref{fig:figure5}, we consider two electrons ($i=2$, $3$) sequentially ejected~\cite{Fletcher2013,ubbelohde2015partitioning} from source S2 and one electron $i=1$ from source S1. When $U_{\rm BS}$ is set around the energy of electron $i=1$, $dI_{\rm D2}/dU_{\rm BS}$ [positive/red feature in Fig.~\ref{fig:figure5}(b)] exhibits two features separated by a time difference $\Delta \bar{t} = 35 \pm 6$ ps. This is the relative injection time of electrons $i=2,3$ from source S2 which are ejected sequentially~\cite{Fletcher2013}. When the beam splitter height $U_{\rm BS}$ varies around the energy of electron $i=2$ or $3$ [negative/blue feature in Fig.~\ref{fig:figure5}(b)] only one feature appears at $\Delta \bar{t} = 0$ or 35 ps. Collision events between the electrons $i=1$ and $2$ do not affect the electron $i=3$ due to the large time difference between the electrons $i=2,3$. 

The temporal size $\tau_{ee} \simeq 6$ ps of the features is set by the temporal wavepacket widths $\sigma_{ti}$ of the electrons and the curvatures of the splitter which determines $\tau_{\omega} \simeq 2$~ps here. In barrier-based methods~\cite{Kataoka2016, Fletcher2013} the temporal resolution is set by the bandwidth of the RF lines controlling the probe barrier~\cite{fletcher2019continuous} whereas the resolution here is set by the emission distribution~\cite{fletcher2019continuous} and beam splitter which suggest a further improvement of the resolution without additional high frequency signals. This multi-electron case also shows the variability of interaction coupling with distance. Interactions between electrons 1 and 3 results ($\Delta \bar{E} \simeq -9.5$~meV) give $\delta \bar{U}_i$ $\simeq 60$~\% smaller than that between electrons 1 and 2 ($\Delta \bar{E} \simeq -5$~meV) consistent with the predictions of Fig.~\ref{fig:figure4}(b).

{\it Conclusion.--} In conclusion, we have show a scheme for sensing single ballistic electrons on picosecond timescales utilizing long-range Coulomb repulsion between electrons. We have also shown a model framework for reading out the geometrical parameters of the beam splitter~\cite{GeierPRB2020} and identifying the strength of the Coulomb interaction. Optimizing the splitter will be valuable for quantifying the error in tomographic measurements~\cite{fletcher2019continuous}, achieving energy-independent beam splitter for electron interferometry and single-qubit operations~\cite{Bauerle_2018}, designing two-qubit operations~\cite{Bauerle_2018}, and probing disorder inside the splitter~\cite{AonoPhysRevB.102.045305}.

{\it Acknowledgement.--} This work was supported by the UK government’s Department for Business, Energy and Industrial Strategy and from the Joint Research Projects 17FUN04 SEQUOIA from the European Metrology Programme for Innovation and Research (EMPIR) co-financed by the participating states and from the European Union’s Horizon 2020 research and innovation programme. It was also supported by Korea NRF (SRC Center for Quantum Coherence in Condensed Matter, Grants No. RS-2023-00207732, No. 2023R1A2C2003430, and No. 2021R1A2C3012612).

\bibliographystyle{apsrev4-2}

\clearpage
\title{Supplementary material: Coulomb sensing of single ballistic electrons}
\maketitle
\setcounter{equation}{0}
\setcounter{figure}{0}
\setcounter{table}{0}
\setcounter{page}{1}
\setcounter{secnumdepth}{3}

\makeatletter
\renewcommand*{\thesection}{S\arabic{section}}
\renewcommand*{\thesubsection}{\thesection.\arabic{subsection}}
\renewcommand*{\p@subsection}{}
\renewcommand*{\thesubsubsection}{\thesubsection.\arabic{subsubsection}}
\renewcommand*{\p@subsubsection}{}
\renewcommand{\bibnumfmt}[1]{[S#1]}
\renewcommand{\citenumfont}[1]{S#1}
\renewcommand{\theequation}{S\arabic{equation}}
\renewcommand{\thefigure}{S\arabic{figure}}
\renewcommand{\theHfigure}{S\arabic{figure}}
\makeatother
\onecolumngrid

\section{Experimental details}

\subsection{Variation of mean injection energy with injection phase}
\label{sec:fluctuation}

\begin{figure}[h]
\centering
\includegraphics[width=0.4\columnwidth]{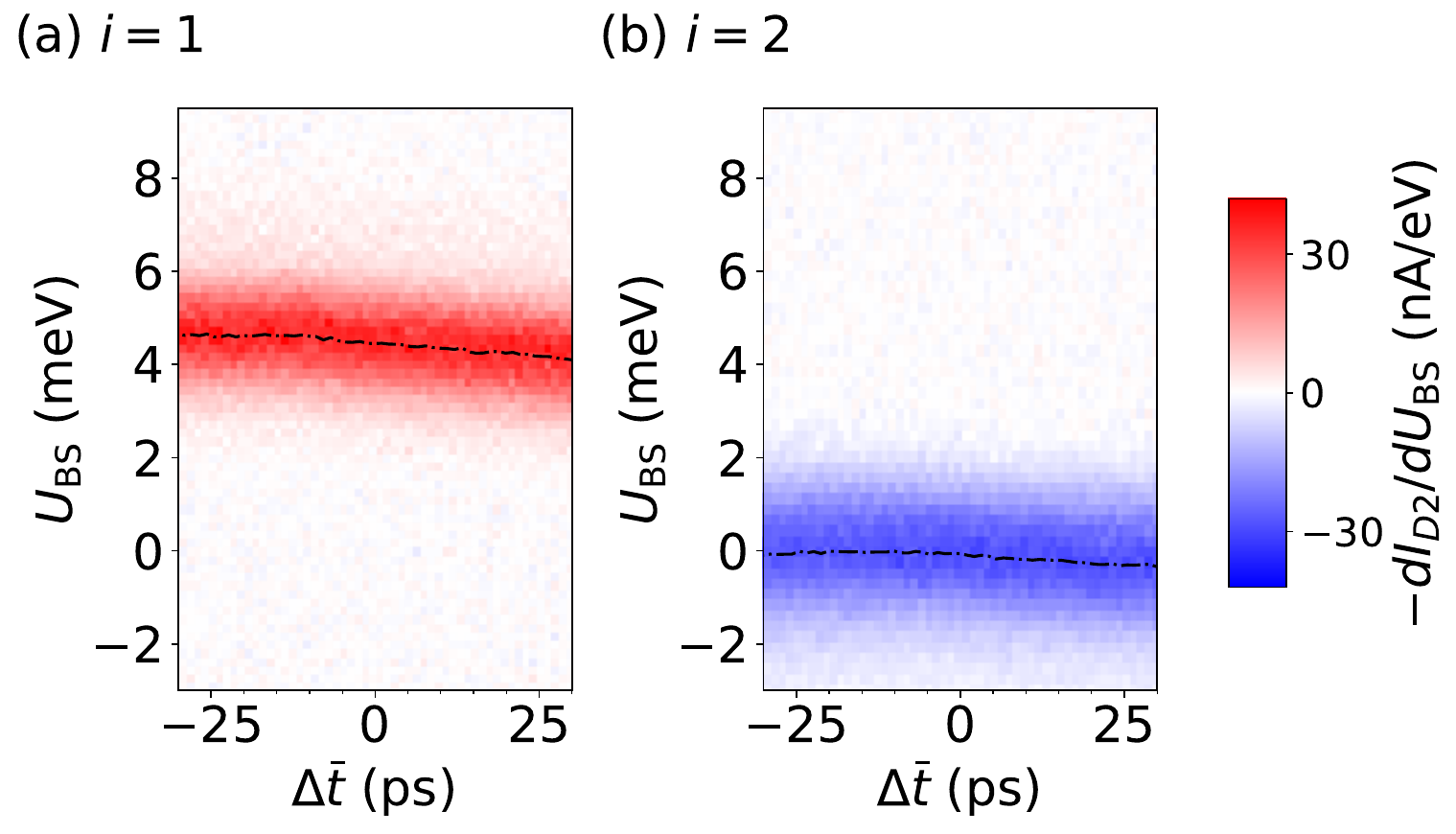}
\caption{$dI_{\rm{D2}}/dU_{\rm BS}$ with respect to $\Delta \bar{t}$ and $U_{\rm BS}$ when (a) the source S1 turns on while turning off the source S2 and (b) S2 only turns on while turning off S1. $\bar{E}_{i}$ computed by the Gaussian fitting of $dI_{\rm{D2}}/dU_{\rm BS}$ over $U_{\rm BS}$ is shown as the dashed-dotted line.}
\label{fig:dIdEB_ref}
\end{figure}

As shown in the paper, the apparent energy $\bar{U}_i (\Delta \bar{t})$ of electron $i$ at the beam splitter varies due to the Coulomb interaction near $\Delta \bar{t}\simeq 0$. Over a broader range of $\Delta \bar{t}$ there is a small additional variation due to a cross-talk effect between the pump driving gates and the beam splitter. This can be systematically isolated and subtracted as follows: All gate AC and DC sources are set the operating point for single electron emission and synchronising partitioning. A small shift to one of the pump DC gate voltages preserves the same cross-talk signature but deactivates that pump. The Coulomb interaction effects disappears from the partitioning signal of remaining active pump but the background variation in the measured mean injection energy $\bar{E}_i(\Delta \bar{t})$ (dashed-dotted lines) remains. This is then repeated for the other pump; see the measured transconductance in Fig.~\ref{fig:dIdEB_ref} compared to Fig.~3(a) in the main text. Using this observation, we compensate the variation of the injection energy by subtracting the term $\bar{E}_i(\Delta \bar{t}) -  \bar{E}_i(\Delta \bar{t} \to \infty)$ in obtaining $\delta \bar{U}_i$ [Eq.~(2) in the main text] from experimental data, i.e.,
\begin{equation}
\delta \bar{U}_i (\Delta \bar{t}) = \bar{U}_i (\Delta \bar{t}) -  \bar{U}_i (\Delta \bar{t} \to \infty) - [ \bar{E}_i(\Delta \bar{t}) -  \bar{E}_i(\Delta \bar{t} \to \infty)].
\label{eq:deltabarUn_compensated}
\end{equation}
We note that the strength of the variation is much smaller than $1$ meV in the time window $\tau_{ee}$ of the relative mean injection time $\Delta \bar{t}$, over which the Coulomb interaction contributes significantly. Hence, we can use a single value of the relative injection energy $\Delta \bar{E}$ in the main text.

\subsection{Wigner distribution of incident electrons}
\label{sec:Wigner}

\begin{figure}[h]
\centering
\includegraphics[width=0.4\columnwidth]{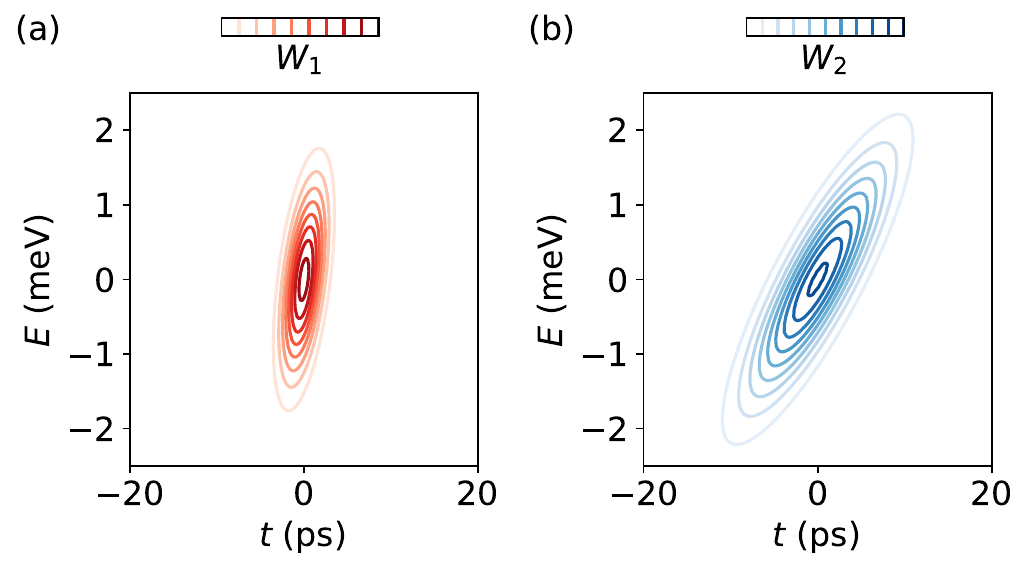}
\caption{Wigner distribution of the electron coming from the sources (a) S1 and (b) S2. See the parameters in the text.}
\label{fig:Wigner_dist}
\end{figure}

Here we introduce parameters of Wigner distribution of incident electrons. Using a time-modulated barrier in the path of incident electrons, one can reconstruct their Wigner distribution~\cite{fletcher2019continuous-2, fletcher2023time-2}. We use results in Ref.~\cite{fletcher2023time-2} (having the same condition except for the mean injection energy), where they assume the bivariate Gaussian function $W_i$ with parameters of time width $\sigma_t$, energy width $\sigma_E$, and energy-time correlation coefficient $r$
\begin{equation}
W_i(\sigma_t, \sigma_E, r) = \dfrac{1}{2 \pi \sigma_t \sigma_E \sqrt{1-r^2}} \exp \bigg[ - \dfrac{1}{2(1-r^2)} \bigg( \dfrac{E^2}{\sigma_E^2} + \dfrac{t^2}{\sigma_t^2} - \dfrac{2 r E t}{\sigma_E \sigma_t} \bigg) \bigg],
\end{equation}
and find the parameters using the least square method.  Here, $E$ and $t$ are the injection energy and time (equivalent to the arrival time at the detector) of the injected electron, relative to the mean value, respectively.  Results are obtained as 
$\sigma_{t1} = 1.7 \pm 0.5$ ps, $\sigma_{E1} = 0.85 \pm 0.2$ meV, $r_1 \simeq 0.5$
for the electron coming from the left source S1, 
and $\sigma_{t2} = 5.2 \pm 1$ ps, $\sigma_{E2} = 1.05 \pm 0.2$ meV,  $r_2 \simeq 0.85$ for the electron from the right source S2~\cite{fletcher2023time-2}.
See Fig.~\ref{fig:Wigner_dist} for graphical representation of the Wigner distribution.

\subsection{Quantification of errors in $\delta \bar{U}_i$}
\label{sec:error}

\begin{figure}[h]
\centering
\includegraphics[width=0.32\columnwidth]{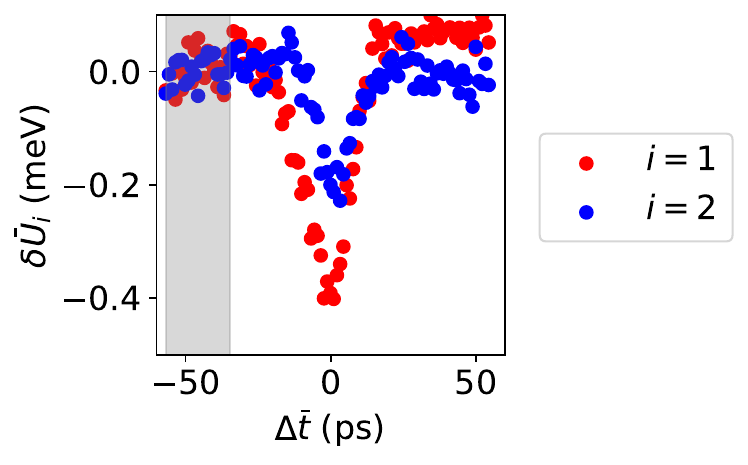}
\caption{Experimental values of $\delta \bar{U}_i$ as a function of $\Delta \bar{t}$ for $\Delta \bar{E}= -6$ meV. This is the extended data of Fig. 3(c) in the main text. The gray-shaded region indicates where $\sigma_u$ is computed.}
\label{fig:error}
\end{figure}

We observe fluctuations in experimental value of $\delta \bar{U}_i$ with respect to $\Delta \bar{t}$ even in the non-interacting region at which $\Delta \bar{t} \to \pm \infty$ (see Fig.~\ref{fig:error}). To quantify this error (denoted as $\sigma_u$) by the fluctuations, we calculate the standard deviation (uncertainty) of $\delta \bar{U}_i$ over $40$ data points within the gray-shaded non-interacting region shown in Fig.~\ref{fig:error}. The error is found to be $\sigma_u \simeq 0.03$ meV. This value is used to determine the error bars in Fig.~4b of the main text and to compute the reduced chi-square $\chi^2_{\nu}$ presented in Fig.~\ref{fig:chi_square}.

\section{Theoretical method}
\label{sec:theoretical}

We provide the theoretical method used to compute maps of detector current in the main text. The method is based on classical guiding center trajectories connecting the electron sources and the detectors, and on ensembles describing incident wave packets (namely, the initial points of the trajectories). We then provide analytical equations used for quantitative study of the potential barrier geometry and the Coulomb interaction in the main text.

\subsection{Detector current}
\label{sec:detector}

We introduce the theoretical method~\cite{fletcher2023time-2} used for calculating a detector current, after discussing its validity.
The method is valid in the parameter regime of our experiment where $\Delta_b$ is much smaller than the energy width $\sigma_{Ei}$ of the incident wave packets.
Here $\Delta_b$ is the energy window over which the transmission probability of a plane wave through the potential barrier changes from $0$ to $1$.
In this regime, the portion of the incident wave packets undergoing quantum tunneling through the barrier is small enough so that classical trajectories and their initial ensemble can describe observables~\cite{park2023measuring-2}.
In our experiment, we estimate that the tunneling energy window $\Delta_b \simeq 0.1$ meV is sufficiently smaller than the packet energy widths $\sigma_{Ei} \simeq 1$ meV. The estimation of $\hbar/\Delta_b$ can be done from the temporal width $\tau_{ee}$ over which the Coulomb interaction is significant in $dI_{\rm D2}/dU_{\rm BS}$, as discussed below.

The guiding center trajectory of electrons is determined by the  $\vec{E} \times \vec{B}$  drift motion along equipotential lines,
\begin{equation}
\vec{v}_i \equiv \frac{d \vec{r}_i}{dt} = \frac{1}{e|B|^2} \vec{\nabla}_i \mathcal{U}_{\rm{tot}} \times \vec{B},
\label{eq:velocity}
\end{equation}
where $\vec{r}_i(t) = (x_i(t), y_i(t))$ is the trajectory of an electron starting from an initial position $(x_{0,i}, y_{0,i})$,  $\vec{B} = -|B| \hat{z} $ is the magnetic field applied perpendicular to the two-dimensional electron gas, and $e$ is the elementary charge.  The index $i=1,2$ stands for an electron generated by the source $i$. $\mathcal{U}_{\text{tot}}$ is the potential determined by the electrostatic potentials $\mathcal{U}_{\text{2D}}$ of the beam splitter  and the inter-particle Coulomb potential $\mathcal{U}_{\text{ee}}(\vec{r}_1, \vec{r}_2)$ between two incident electrons,
\begin{equation}
\mathcal{U}_{\textrm{tot}}(\vec{r}_1,\vec{r}_2 ) = \mathcal{U}_{\textrm{2D}}(\vec{r}_1) + \mathcal{U}_{\textrm{2D}}(\vec{r}_2) + \mathcal{U}_{ee}(\vec{r}_1, \vec{r}_2).
\label{eq:Utot}
\end{equation} 
We model $\mathcal{U}_{\text{2D}}$ as the saddle constriction potential~\cite{fertig1987transmission-2} whose curvatures along $x$ and $y$ are determined by $\omega_x$ and $\omega_y$,
\begin{equation}
\mathcal{U}_{\textrm{2D}}(x, y) = U_{\rm BS} - \frac{1}{2} m^* \omega_x^2 x^2 + \frac{1}{2} m^* \omega_y^2 y^2,
\label{eq:U_2D}
\end{equation} 
where $m^*$ is the effective electron mass in GaAs and $U_{\rm BS}$ is the beam splitter height at the saddle point. The tunneling energy window $\Delta_b$ of the saddle potential is $\Delta_b = \hbar \omega_x  \omega_y / (2 \sqrt{3} \omega_c)$, where $\omega_c$ is the cyclotron frequency, $\omega_c \simeq 26.25 $ ps$^{-1}$ for $|B| =10$ T. The Coulomb potential between the two electrons is described by
\begin{equation}
\mathcal{U}_{\textrm{ee}}(\vec{r}_1, \vec{r}_2) = \beta \left[ \dfrac{1}{|\vec{r}_1 - \vec{r}_2|} - \dfrac{1}{[ (\vec{r}_1 - \vec{r}_2)^2 + (2 d_{\textrm{2DEG}} )^2 ]^{1/2}} \right],
\label{eq:U_ee}
\end{equation}
where $\beta$ is the interaction strength. The second term of Eq.~\eqref{eq:U_ee} describes the effect of the partial screening~\cite{skinner2010anomalously-2} by the surface gate, which significantly reduces the Coulomb potential when the distance between the electrons is much larger than the distance $d_{\text{2DEG}} \simeq 90$~nm between the two-dimensional electron gas and the surface gate. The presence of the partial screening also justifies the saddle potential model since the Coulomb interaction is negligible outside $d_{\rm{2DEG}}$. The interaction strength in GaAs is found as $\beta = e^2 / (4 \pi \epsilon_0 \epsilon_r) \simeq 112 \text{ meV$*$nm}$, where $\epsilon_0$ is the vacuum permittivity, $\epsilon_r = 12.9$ is the relative permittivity of GaAs, and $e$ is the elementary charge.

The ensemble of the initial points of the classical trajectories, describing an incident wave packet of the electron~$i$, is given by the Wigner distribution $W_i (E_i, t_i)$ in energy-time space, which is obtained by the tomographic measurement~\cite{fletcher2023time-2}. See Sec.~\ref{sec:Wigner} for the parameters of the Wigner distributions. Here, an initial point is located along the equipotential line of $E_i = \mathcal{U}_{\text{2D}}(x_{0,i}, y_{0,i})$ at the injection time $t_i$.

We compute the current $I_{\rm D2}$ at the detector D2 ($x$ coordinate of detector D2 is positive) using this method,
\begin{equation}
I_{\rm D2} = (ef) \int^{\infty}_{-\infty} d E_1 \int^{\infty}_{-\infty} d t_{1} \int^{\infty}_{-\infty} d E_2 \int^{\infty}_{-\infty} d t_{2} \, W_1(E_1, t_{1}) W_2(E_2, t_{2}) \lim_{t \to \infty}   \big[\Theta \big( x_{1,\text{cl}}(t) \big) + \Theta \big( x_{2,\text{cl}}(t) \big) \big],
\label{eq:current}
\end{equation}
where $\Theta(x)$ is $1$ for $x>0$ and $0$ otherwise, $f$ is a pumping frequency. The positions $\vec{r}_{1,\text{cl}}(t)$ and $\vec{r}_{2,\text{cl}}(t)$ of the electrons are obtained by solving the classical equation of motion [Eq.~\eqref{eq:velocity}] with initial positions determined by injection energy and time, $(E_1, t_1)$ and $(E_2, t_2)$, respectively. We note that the current conservation, $I_{\rm D1} + I_{\rm D2} = 2ef$, is satisfied in the computation.

Equation~\eqref{eq:current} can be equivalently written as
\begin{align}
\begin{split}
I_{\rm D2} = (ef) \int^{\infty}_{-\infty} d E_1 & \int^{\infty}_{-\infty} d t_{1} \int^{\infty}_{-\infty} d E_2 \int^{\infty}_{-\infty} d t_{2} \, W_1(E_1, t_{1}) W_2(E_2, t_{2}) \\
& \times \big[ \Theta [E_1 - U_{\rm BS} + \delta U_1(\Delta t, \Delta E) ] + \Theta [-E_2 + U_{\rm BS} - \delta U_2(\Delta t, \Delta E)] \big],
\label{eq:current2}
\end{split}
\end{align}
where $\delta U_i$ is the difference of the classical barrier height threshold (to block the electron $i$) between presence and absence of the Coulomb interaction when the classical trajectories of the two electrons have the relative energy $\Delta E \equiv E_2 - E_1$ and the relative injection time $\Delta t \equiv t_2  - t_1$.

\begin{figure}
\centering
\includegraphics[width=0.45\columnwidth]{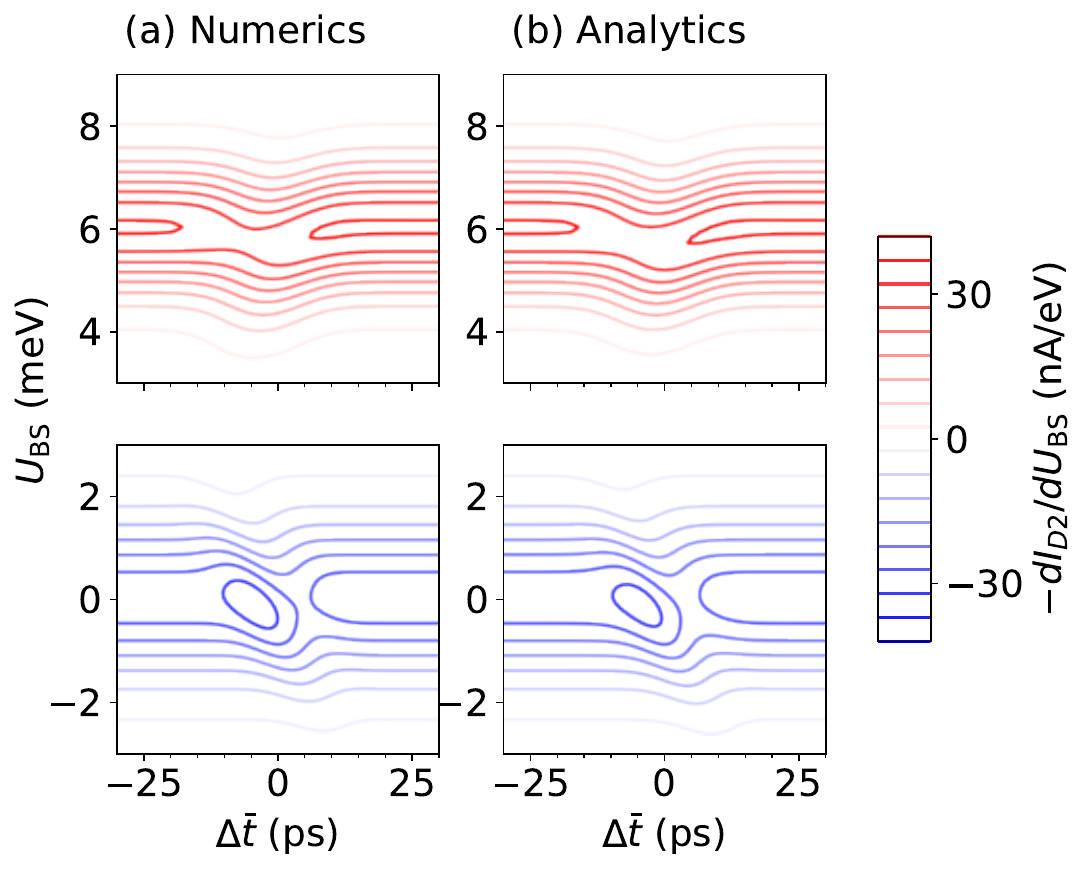}
\caption{$dI_{\rm D2}/dU_{\rm BS}$ computed using (a) exact numerical computation and (b) analytical equations of Eqs.~\eqref{eq:dIdEb1}--\eqref{eq:dIdEb2} with Eq.~\eqref{eq:deltaEn_analytic}. They are drawn with $\Delta \bar{E} = -6$ meV, $\omega_x = 5.8$ ps$^{-1}$, $\omega_y = 2.2$ ps$^{-1}$, and $\beta = 95$ meV$*$nm. In the numerical computation, the current $I_{\rm D2}$ of Eq.~\eqref{eq:current} is calculated by computing classical trajectories numerically.}
\label{fig:dIdEB}
\end{figure}

\subsection{Analytical equations for the regime of large relative injection energy}

For quantitative analysis, we consider the regime where the relative mean injection energy $|\Delta \bar{E}|$ between the two electrons is much larger than the energy widths $\sigma_{Ei}$ of incident wave packets and the difference $|\delta U_i|$ of the classical barrier height threshold between the presence and absence of the Coulomb interaction (e.g., around $\Delta \bar{E} = - 6$ meV in our experiments).
In this regime, the electrons almost independently contribute to $dI_{\rm D2}/dU_{\rm BS}$ [see Fig.~\ref{fig:dIdEB}(a) for $\Delta \bar{E} = -6$ meV], and the difference $\delta \bar{U}_i$ of the threshold barrier height by the Coulomb interaction is well defined and computed perturbatively.
We derive an analytical equation of $\delta \bar{U}_i$ and analyze the dependence of $\delta \bar{U}_i$ on the saddle potential parameters and the Coulomb interaction strength.

\subsubsection{Analytical equation for $\delta \bar{U}_i$ }

We first consider the transconductance $dI_{\rm D2}/dU_{\rm BS}$. In the regime of large relative mean injection energy $|\Delta \bar{E}|$ between the two electrons (i.e., $|\Delta \bar{E}| \gg \sigma_{Ei}$), $\Delta E$ in the expression of $\delta U_i (\Delta t, \Delta E)$ [see Eq.~\eqref{eq:current2}] can be substituted by $\Delta \bar{E}$ within the energy window of incident wave packets. Then $dI_{\rm D2}/dU_{\rm BS}$ is found from Eq.~\eqref{eq:current2}, 
\begin{align}
\dfrac{dI_{\rm D2}}{d U_{\rm BS}} & \simeq - (ef) \int^{\infty}_{-\infty} dt_{1} \int^{\infty}_{-\infty} dt_{2} \, \rho_2(t_2) W_1 (E_1, t_{1})  |_{E_1 = U_{\rm BS} - \delta U_{1} (\Delta t, \Delta \bar{E})} \quad \text{for } U_{\rm BS} \simeq \bar{E}_1, 
\label{eq:dIdEb1} \\
\dfrac{dI_{\rm D2}}{d U_{\rm BS}} & \simeq (ef) \int^{\infty}_{-\infty} dt_{1} \int^{\infty}_{-\infty} dt_{2}  \, \rho_1 (t_{1}) W_2(E_2, t_2) |_{E_2 = U_{\rm BS} - \delta U_{2} (\Delta t, \Delta \bar{E})} \quad \text{for } U_{\rm BS} \simeq \bar{E}_2,
\label{eq:dIdEb2}
\end{align}
where $\Delta t \equiv t_2 - t_1$ and $\rho_i \equiv \smallint^{\infty}_{-\infty} dE \, W_i$ is the injection time distribution of electron $i$.  For given $\Delta \bar{E}$, an analytic expression of $\delta U_i$ is found in the next subsection. Using Eqs.~\eqref{eq:dIdEb1} and \eqref{eq:dIdEb2}, and the definition of $\delta \bar{U}_i$ of Eq.~(1) in the main text,  $\delta \bar{U}_i$ is evaluated as the weighted-average of $\delta U_i$ with the time distributions,
\begin{align}
\delta \bar{U}_{i}(\Delta \bar{t}, \Delta \bar{E}) & \simeq \int^{\infty}_{-\infty} dt_1 \int^{\infty}_{-\infty} dt_2 \, \rho_1(t_1) \rho_2(t_2) \delta U_{i}(\Delta t, \Delta \bar{E}) \\
& = \int^{\infty}_{-\infty} d \Delta t \, \tilde{\rho}(\Delta t; \Delta \bar{t}) \delta U_{i} (\Delta t, \Delta \bar{E}),
\label{eq:barEb}
\end{align}
where $\tilde{\rho}(t; \Delta \bar{t}) \equiv \int^{\infty}_{-\infty} dt' \, \rho_1(t') \rho_2(t + t')$. The mean value and standard deviation of $\tilde{\rho}$ are $\Delta \bar{t} \equiv \bar{t}_2 - \bar{t}_1$ and $\sqrt{\sigma_{t1}^2 + \sigma_{t2}^2}$, respectively. $\bar{t}_i$ and  $\sigma_{ti}$ are the mean value and the temporal width of the injection time distribution $\rho_i$ of the electron $i$, respectively. 
We note that $\delta \bar{U}_i$ and $\delta U_i$ depend on $\Delta \bar{E}$; in the main text, this dependence is omitted for simplicity. We also note that Eq.~\eqref{eq:barEb} has a convolutional form with respect to $\Delta \bar{t}$, since $\tilde{\rho}(t;\Delta \bar{t}) = \tilde{\rho}(t - \Delta \bar{t}; 0)$.

We found that the approximated $dI_{\rm D2}/dU_{\rm BS}$ of Eqs.~\eqref{eq:dIdEb1}--\eqref{eq:dIdEb2}, with analytical $\delta U_i$ [Eq.~\eqref{eq:deltaEn_analytic}] shown later, matches the exact numerical computation of $dI_{\rm D2}/dU_{\rm BS}$ under our experimental condition of $\Delta \bar{E} = -6$ meV [Fig.~\ref{fig:dIdEB}]. This justifies the use of the analytical equations in the main text for extracting system parameters.

\subsubsection{Analytic expression of $\delta U_i$}

\begin{figure}[t]
\centering
\includegraphics[width=0.45\columnwidth]{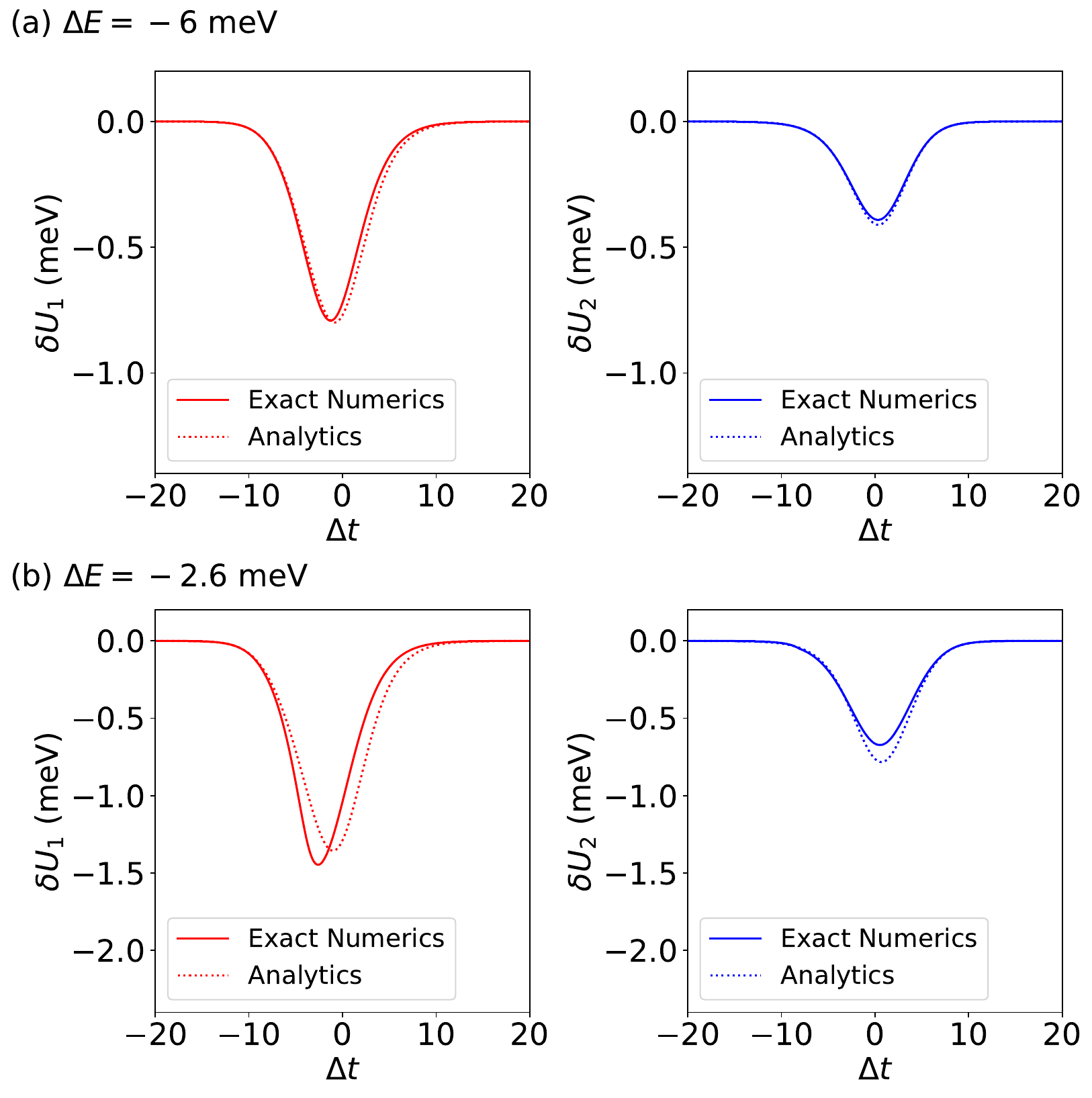}
\caption{$\delta U_1$ and $\delta U_2$ as a function of $\Delta t$ at (a) $\Delta E = -6$ meV and (b) $\Delta E = -2.6$ meV. 
The dotted lines show the perturbative result obtained from Eq.~\eqref{eq:deltaEn_analytic},
and the solid lines show the exact result obtained from numerical computation of classical trajectories. They are drawn with $\omega_x = 5.8$ ps$^{-1}$, $\omega_y = 2.2$ ps$^{-1}$, and $\beta = 95$ meV$*$nm. }
\label{fig:deltaE_comparison}
\end{figure}

We derive an analytic expression of the difference $\delta U_i$ of the classical height threshold between the presence and absence of the Coulomb interaction in the regime of large $|\Delta \bar{E}|$, where a kinetic energy variation of electrons by the Coulomb interaction is small enough to be treated perturbatively. In the derivation, we use that $\delta U_i$ is equivalent to the kinetic energy loss for electron $i$ to be eventually trapped in the saddle point at $t \to \infty$. The kinetic energy $\mathcal{U}_{\rm 2D}$ of each classical electron $i$ varies over time as
\begin{align}
\dfrac{d}{dt} \mathcal{U}_{\rm 2D} \big(\vec{r}_i(t) \big) & = \vec{v}_i(t) \cdot \vec{\nabla}_i \mathcal{U}_{\rm 2D} \big(\vec{r}_i(t) \big) \\ & = - \vec{v}_i(t) \cdot \vec{\nabla}_i \mathcal{U}_{ee} \big( \vec{r}_1(t), \vec{r}_2(t) \big),
\label{eq:Uee}
\end{align}
where $\vec{v}_i$ is the velocity of the electron $i$ [Eq.~\eqref{eq:velocity}] and we use the conservation of the total energy $\mathcal{U}_{\text{tot}}$ of Eq.~\eqref{eq:Utot} in the second equality. We note that the classical result of Eq.~\eqref{eq:Uee} is consistent with previous results~\cite{ryu2022partition-2} of kinetic energy variations of two interacting electrons propagating toward a barrier: For two co-propagating electrons, the electron arriving earlier at the barrier gains an energy due to the interaction while the other loses the energy. For two counter-propagating electrons moving to the barrier in the opposite direction to each other~\cite{fletcher2023time-2}, the electron later arriving at the barrier loses more energy than the other.

Integrating Eq.~\eqref{eq:Uee} over time and using the classical $\vec{E} \times \vec{B}$ drift trajectories $\vec{r}_i (t')$ and velocities $\vec{v}_i (t')$ of the electron $i$ obtained in the absence of the Coulomb interaction [obtained by solving Eq.~\eqref{eq:velocity} with $\mathcal{U}_{\rm tot}$ replaced by $\mathcal{U}_{\rm 2D}$, see Sec.~\ref{sec:detail} in details], $\delta U_i$ is evaluated as
\begin{align}
	\delta U_{i}(\Delta t, \Delta E) & \simeq - \int^{\infty}_{-\infty} dt' \, \vec{v}_i(t') \cdot \vec{\nabla}_i \mathcal{U}_{\text{ee}} \big( \vec{r}_1(t'), \vec{r}_2(t') \big).
	\label{eq:DeltaEbn}
\end{align}
In the evaluation of $\delta U_1$, $\vec{r}_1(t')$ and $\vec{r}_2(t')$ are the trajectories of the electrons at time $t'$, obtained with the initial conditions of $E_1 \to U_{\rm BS}$, $E_2 \to U_{\rm BS} + \Delta E$, and time delay $\Delta t$ between the two electrons. 
In the calculation of  $\delta U_2$, $\vec{r}_1(t')$ and $\vec{r}_2(t')$ are obtained with the initial conditions of $E_1 \to U_{\rm BS} - \Delta E$, $E_2 \to U_{\rm BS}$, and $\Delta t$.
The integral in Eq.~\eqref{eq:DeltaEbn} is analytically found as [see Eqs.~\eqref{eq:deltaE1_integral_form}--\eqref{eq:deltaE2_integral_form} in Sec.~\ref{sec:detail} for the derivation]
\begin{equation}
\delta U_{i}(\Delta t, \Delta E) = -\dfrac{\beta}{\sqrt{\Delta E}}  f_i(\Delta t; \omega) g_i(\omega_x, \omega_y) + \textrm{screening term}.
\label{eq:deltaEn_analytic}
\end{equation}
The first term of Eq.~\eqref{eq:deltaEn_analytic} comes from the unscreened Coulomb potential 
$\beta/r$ in $\mathcal{U}_{ee}$ [see the first term of Eq.~\eqref{eq:U_ee}], while the second term describes the effect of the screening term $\beta/\sqrt{r^2+(2d_{\rm{2DEG}})^2}$ in $\mathcal{U}_{ee}$ [the second term of Eq.~\eqref{eq:U_ee}]. This screening term is also obtained analytically (see Sec.~\ref{sec:detail}). We numerically confirm that Eq.~\eqref{eq:deltaEn_analytic} (including the screening term) is in good agreements with exact numerical computations at $\Delta E = -6$ meV with parameters in the main text [see Fig.~\ref{fig:deltaE_comparison}(a)]. The non-perturbative effect of $\delta U_i$ [the deviation of $\delta U_i$ from the approximated equation \eqref{eq:deltaEn_analytic}] begins to play a role at the injection energy difference of $\Delta E = - 2.6$ meV [see Fig.~\ref{fig:deltaE_comparison}(b)], the smallest energy difference in our experiment. The detailed expression of the first term in Eq.~\eqref{eq:deltaEn_analytic} follows
\begin{equation}
f_1(\Delta t; \omega) = \dfrac{\exp[-\omega \Delta t]}{(1 + \exp[-\omega \Delta t])^{3/2}},
\label{eq:f1}
\end{equation}
\begin{equation}
f_2(\Delta t; \omega) = \dfrac{\exp[\omega \Delta t]}{(1 + \exp[\omega \Delta t])^{3/2}},
\label{eq:f2}
\end{equation}
\begin{equation}
g_1(\omega_x, \omega_y) = \dfrac{1}{\sqrt{2}} \sqrt{ m^{*} } \dfrac{\omega_x \omega_y}{\sqrt{\omega_x^2 + \omega_y^2}}  \int^{0}_{-\infty} \dfrac{dx}{\sqrt{1 + x^4 \mp 2 x^2 \cos(2 \theta)}},
\label{eq:g1}
\end{equation}
\begin{equation}
g_2(\omega_x, \omega_y) =  \dfrac{1}{\sqrt{2}} \sqrt{ m^{*} } \dfrac{\omega_x \omega_y}{\sqrt{\omega_x^2 + \omega_y^2}} \int^{0}_{-\infty} \dfrac{dx}{\sqrt{1 + x^4 \pm 2 x^2 \cos(2 \theta)}},
\label{eq:g2}
\end{equation}
where the upper signs in the $\pm$ and $\mp$ factors in $g_1$ and $g_2$ correspond to $\Delta E<0$, while the lower signs correspond to $\Delta E>0$. Here, $\theta \equiv \text{arctan}(\omega_y/\omega_x)$ and $\omega \equiv \omega_x \omega_y/\omega_c$. The signs of $\exp[\pm \omega \Delta t]$ in Eqs.~\eqref{eq:f1}--\eqref{eq:f2} indicate that the Coulomb interaction effects depend on which one between the two electrons earlier arrives at the barrier. In our convention, when $\Delta t>0$, the electron coming from the right source S2 arrives at the barrier later than the other electron from the left source S1. The detailed expression is discussed in the next subsection.

\begin{figure} [t]
	\centering
	\includegraphics[width=0.62\columnwidth]{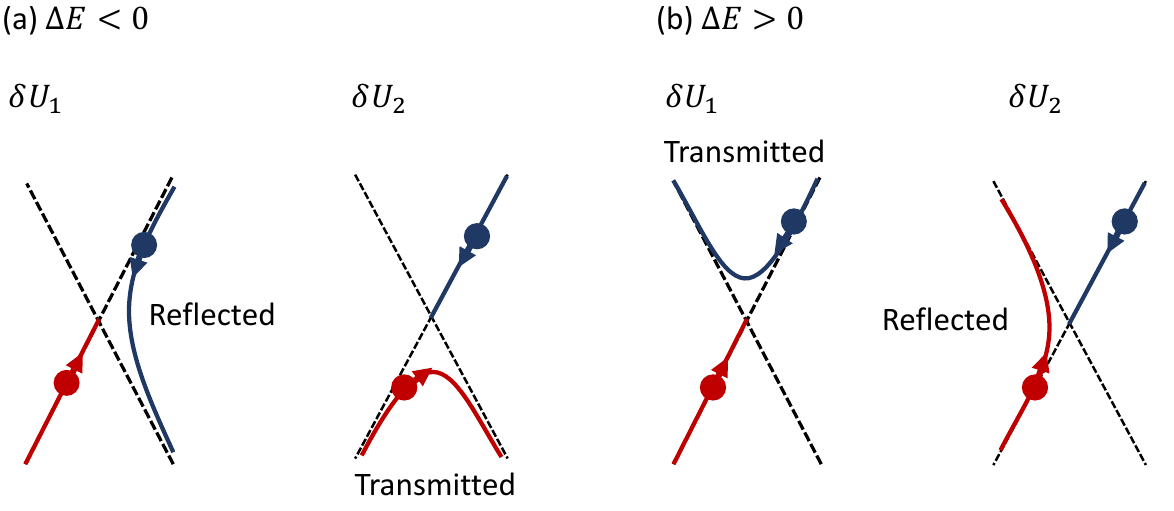}
	\caption{The electron trajectories (red and blue solid lines) for (a) $\Delta E<0$ and (b) $\Delta E>0$ with $\omega_x > \omega_y$. The left panel represents the trajectories affecting $\delta U_1$, and the right panel represents those affecting $\delta U_2$. The trajectories have the inversion symmetry for the different polarity of $\Delta E$.} 
	\label{fig:polarity}
\end{figure}

\begin{figure}[t]
\centering
\includegraphics[width=0.4\columnwidth]{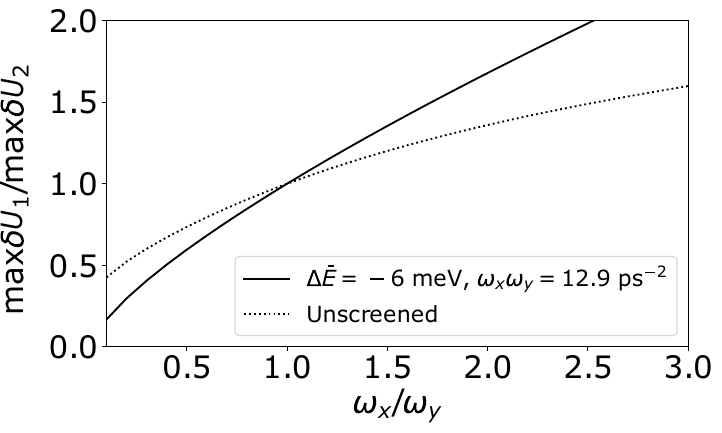}
\caption{$\max{\delta U_1}/\max{\delta U_2}$ as a function of $\omega_x / \omega_y$ and $\Delta E<0$. The case where only the unscreened potential part is taken into account in Eq.~\eqref{eq:deltaEn_analytic} is shown by the dotted curve. The case where both the unscreened part and the screening term are taken into account in Eq.~\eqref{eq:deltaEn_analytic} is shown by the solid curve. See Fig.~\ref{fig:polarity}(a) for electrons trajectory affecting $\delta U_1$ and $\delta U_2$.}
\label{fig:deltaE_ratio}
\end{figure}

\subsubsection{Dependence of $\delta \bar{U}_i$ on system parameters}

We analyze the dependence of the difference $\delta \bar{U}_i$ of the threshold barrier height on
the saddle potential parameters $\omega_x$, $\omega_y$, and the strength $\beta$ of the Coulomb repulsion. 

Firstly, from the convolutional form of Eq.~\eqref{eq:barEb} we find that
the temporal width $\tau_{ee}$ of the $\Delta \bar{t}$-dependence of $\delta \bar{U}_i$ follows
\begin{equation}
\tau_{ee} = \sqrt{ \sigma_{t1}^2 + \sigma_{t2}^2 + \tau_{\omega}^2}.
\label{eq:tee}
\end{equation}
$\tau_{\omega}$ is the temporal width of the $\Delta t$-dependence of $\delta U_i$ and $\sigma_{ti}$ is the temporal width of the injection time distribution $\rho_i$ of electron $i$. $\tau_{\omega}$ is  order of  $1/\omega$ (here $ \omega  = \omega_x \omega_y/\omega_c$), the characteristic time scale of the saddle potential. This is attributed to the fact that, for $|\Delta t| \gg 1/\omega$, the two incident electrons miss each other, and the system goes to the non-interacting case.  When only the unscreened potential part is taken into account in $\delta U_i$ [the first term in Eq.~\eqref{eq:deltaEn_analytic}], from Eqs.~\eqref{eq:f1}--\eqref{eq:f2}, the temporal width $\tau_{\omega}$ is indeed found as $\tau_{\omega} = (\sqrt{2/3})( \pi /\omega)$. The value $\pi \sqrt{2/3} \simeq 2.5651$ of the proportional factor is specific to the $1/r$ Coulomb interaction potential. For $1/r^2$ and $1/r^3$ inter-particle potentials, this factor becomes lower since the potential changes more rapidly with respect to $r$. When both the unscreened part and the screening term are taken into account in Eq.~\eqref{eq:deltaEn_analytic}, the proportional factor to $1/\omega$ changes slightly when other parameters of the setup vary. However, the order of the proportional factor is still $1$.

Secondly, we consider the ratio $\max{\delta \bar{U}_{1}} /\max{\delta \bar{U}_{2}}$, where the maximum values of $\delta \bar{U}_{i}$  are obtained with varying time $\Delta \bar{t}$. $\max {\delta \bar {U_i}}$ is the benchmark for the strength of the interaction when the electrons reach their point of closest approach. From Eq.~\eqref{eq:barEb} and assuming the Gaussian form of the time distribution $\rho_i$, we obtain
\begin{equation}
\dfrac{\max{\delta \bar{U}_{1}}}{\max{\delta \bar{U}_{2}}} = \dfrac{\max{\delta U_{1}}}{\max{\delta U_{2}}}.
\label{eq:ratio}
\end{equation}
It changes with the anisotropy $\omega_x/\omega_y$ of the saddle potential, attributed to the different turning point of the classical trajectories between the electrons $i$, see Fig.~\ref{fig:polarity} for example of $\omega_x > \omega_y$. Due to the inversion symmetry of our saddle potential geometry, for a given $|\Delta \bar{E}|$, $\max{\delta U_1}/\max{\delta U_2}$ with $\Delta E<0$ is the same as $\max{\delta U_2}/\max{\delta U_1}$ with $\Delta E>0$. When only the unscreened potential part is taken into account in Eq.~\eqref{eq:deltaEn_analytic}, we obtain from Eqs.~\eqref{eq:g1}--\eqref{eq:g2}
\begin{equation}
\dfrac{\max{\delta U_{1}}}{\max{\delta U_{2}}} = \dfrac{\int^{0}_{-\infty} dx \, [1 + x^4 \mp 2 x^2 \cos(2 \theta)]^{-1/2}}{\int^{0}_{-\infty} dx \, [1 + x^4 \pm 2 x^2 \cos(2 \theta)]^{-1/2}},
\end{equation}
where the upper signs in the $\pm$ and $\mp$ factors correspond to $\Delta E<0$, while the lower signs correspond to $\Delta E>0$. For $\Delta E<0$, $\max{\delta U_{1}} /\max{\delta U_{2}}$ becomes $1$ when $\omega_{x} = \omega_y$, goes to $>1$ when $\omega_x > \omega_y$, and $<1$ when $\omega_x < \omega_y$. This behavior happens also when both the unscreened part and the screening term are taken into account in Eq.~\eqref{eq:deltaEn_analytic}, see Fig.~\ref{fig:deltaE_ratio}. In the presence of the screening term, the ratio changes more sensitively with respect to the anisotropy factor $\omega_x/\omega_y$ while satisfying the above inequalities.

Thirdly, the intensity of $\delta \bar{U}_i$ is linearly proportional to the strength $\beta$ of the Coulomb potential in our perturbative regime. From the value of $\max{ \delta \bar{U}_i}$, $\beta$ can be estimated. In summary, $\omega_x \omega_y$, $\omega_x/\omega_y$ and $\beta$ are majorly determined by the width of $\delta \bar{U}_i$, the ratio $\max{\delta \bar{U}_{1}} /\max{\delta \bar{U}_{2}}$, and the intensity of $\delta \bar{U}_i$, respectively. Therefore, by comparing the theory and the experiment for $\delta \bar{U}_i$, we can estimate the values of the parameters of the setup with error bars.

We lastly explore the dependence of $\delta \bar{U}_i$ on the relative mean injection energy $\Delta \bar{E}$. From Eqs.~\eqref{eq:barEb} and \eqref{eq:deltaEn_analytic}, we obtain
\begin{equation}
\max{\delta \bar{U}_i} \propto \dfrac{1}{\sqrt{|\Delta \bar{E}|}} + \text{a correction by the screening term.}
\end{equation}
The first term is attributed to the unscreened Coulomb potential $\beta/r$. The second term by the screening by the surface gate is also polynomial with respect to $\Delta \bar{E}$. Hence, observing the dependence of $\delta \bar{U}_i$ on $\Delta \bar{E}$ reveals the long-range characteristic of the Coulomb interaction.

\subsubsection{Full expression of Eq.~\eqref{eq:deltaEn_analytic}}
\label{sec:detail}

We introduce steps to obtain a full expression of Eq.~\eqref{eq:deltaEn_analytic} from $\delta U_{i}(\Delta t, \Delta E) \simeq - \int^{\infty}_{-\infty} dt' \, \vec{v}_i(t') \cdot \vec{\nabla}_i \mathcal{U}_{\text{ee}} \big( \vec{r}_1(t'), \vec{r}_2(t') \big)$ in Eq.~\eqref{eq:DeltaEbn}.
The non-interacting trajectories $r_1(t)$ and $r_2(t)$ are obtained by solving Eq.~\eqref{eq:velocity} with $\mathcal{U}_{\rm tot}$ replaced by $\mathcal{U}_{\rm 2D}$ of Eq.~\eqref{eq:U_2D}, i.e., $dx/dt = - (\omega_y^2/\omega_c) y$ and $dy/dt = - (\omega_x^2/\omega_c) x$, with initial positions satisfying $\lim_{t \to -\infty} [x_2(t)/x_1(t)] = \lim_{t \to -\infty} [y_2(t)/y_1(t)] = -e^{\omega \Delta t}$, and $\mathcal{U}_{\rm 2D}( \vec{r}_1) = E_1$, $\mathcal{U}_{\rm 2D}(\vec{r}_2) = E_2$. We then simplify the integral in Eq.~\eqref{eq:DeltaEbn} to obtain analytical expressions. Noting that the trajectory of the particle $i$ (having energy $E_i = U_{\rm BS}$) follows a straight line in 2D space [see Figs.~\ref{fig:deltaE_ratio}(b)--(c)], we represent the integral as the line integral form over the non-interacting trajectory $\mathcal{C}_i$ of the electron $i$, and use the change of variable $u =  (-\omega_x x + \omega_y y)/\sqrt{2(\omega_x^2+\omega_y^2)}$ and $s =  (\omega_x x + \omega_y y)/\sqrt{2(\omega_x^2+\omega_y^2)}$ (the trajectory of particle $i$ now follows $s$ coordinate). Then, Eq.~\eqref{eq:DeltaEbn} is 
\begin{align}
\delta U_{i}(\Delta t, \Delta E) & = - \int_{\mathcal{C}_i} d \vec{r}_i \cdot \vec{\nabla}_i \mathcal{U}_{\text{ee}} = -\int_{\mathcal{C}_i} ds_i \partial_{s_i} \mathcal{U}_{ee}.
\label{eq:DeltaEbn_detail1}
\end{align}
By expressing the computed non-interacting trajectories in terms of $s_i$ and plugging the trajectories into the integrand of Eq.~\eqref{eq:DeltaEbn_detail1}, and from the form of $\mathcal{U}_{ee}$ [Eq.~\eqref{eq:U_ee}], $\delta U_i$ is computed,
\begin{equation}
\delta U_1(\Delta t, \Delta E) = h_1 \int^{0}_{-\infty} d s_1 \bigg[  \dfrac{  s_1 + a_1 / s_1 }{ (s_1^2 + b_1^2/s_1^2 + 2 a_1)^{3/2} } - \dfrac{  s_1 + a_1 / s_1 }{ (s_1^2 + b_1^2/s_1^2 + 2 a_1 + c_1)^{3/2} } \bigg]
\label{eq:deltaE1_integral_form}
\end{equation}
for $i=1$, and 
\begin{equation}
\delta U_2(\Delta t, \Delta E)= h_2 \int^{0}_{\infty} d s_2 \bigg[  \dfrac{  s_2 + a_2 / s_2 }{ (s_2^2 + b_2^2/s_2^2 + 2 a_2)^{3/2} } - \dfrac{  s_2 + a_2 / s_2 }{ (s_2^2 + b_2^2/s_2^2 + 2 a_2 + c_2)^{3/2} } \bigg]
\label{eq:deltaE2_integral_form}
\end{equation}
for $i=2$. Here, $a_i$, $b_i$, $c_i$, and $h_i$ follow
\begin{equation}
a_1 = \dfrac{e^{-\omega \Delta t }}{e^{\omega \Delta t}+1} \dfrac{\omega_x^2-\omega_y^2}{(\omega_x^2+ \omega_y^2)^2}  \frac{\Delta E}{m^*}, \qquad a_2 = - \dfrac{e^{\omega \Delta t }}{e^{-\omega \Delta t}+1} \dfrac{\omega_x^2-\omega_y^2}{(\omega_x^2+ \omega_y^2)^2}  \frac{\Delta E}{m^*},
\end{equation}
\begin{equation}
b_1 = \dfrac{e^{-\omega \Delta t }}{e^{\omega \Delta t}+1} \dfrac{1}{\omega_x^2+ \omega_y^2}  \frac{\Delta E}{m^*}, \qquad b_2 = -\dfrac{e^{\omega \Delta t }}{e^{-\omega \Delta t}+1} \dfrac{1}{\omega_x^2+ \omega_y^2}  \frac{\Delta E}{m^*},
\end{equation}
\begin{equation}
c_1 = 8 d_{\rm{2DEG}}^2 \dfrac{1}{(e^{\omega \Delta t}+1)^2} \dfrac{\omega_x^2 \omega_y^2}{(\omega_x^2+ \omega_y^2)^2}, \qquad c_2 = 8 d_{\rm{2DEG}}^2 \dfrac{1}{(e^{-\omega \Delta t}+1)^2} \dfrac{\omega_x^2 \omega_y^2}{(\omega_x^2+ \omega_y^2)^2},
\end{equation}
\begin{equation}
h_1 = \sqrt{2} \beta \dfrac{1}{(e^{\omega \Delta t}+1)^2} \dfrac{\omega_x \omega_y}{\omega_x^2+ \omega_y^2}, \qquad h_2 = \sqrt{2} \beta \dfrac{1}{(e^{-\omega \Delta t}+1)^2} \dfrac{\omega_x \omega_y}{\omega_x^2+ \omega_y^2}.
\end{equation}
The integrals in Eqs~\eqref{eq:deltaE1_integral_form}--\eqref{eq:deltaE2_integral_form} are  computed  and expressed analytically with the commercial software of the Mathematica. The term attributed to the unscreened Coulomb potential $\beta/r$ [the first term in $[\cdots]$ at Eqs.~\eqref{eq:deltaE1_integral_form}--\eqref{eq:deltaE2_integral_form}] is simplified more, see Eq.~\eqref{eq:deltaEn_analytic}.

\section{Comparison between experiment and theory}

\begin{figure}[t]
	\centering
	\includegraphics[width=0.25\columnwidth]{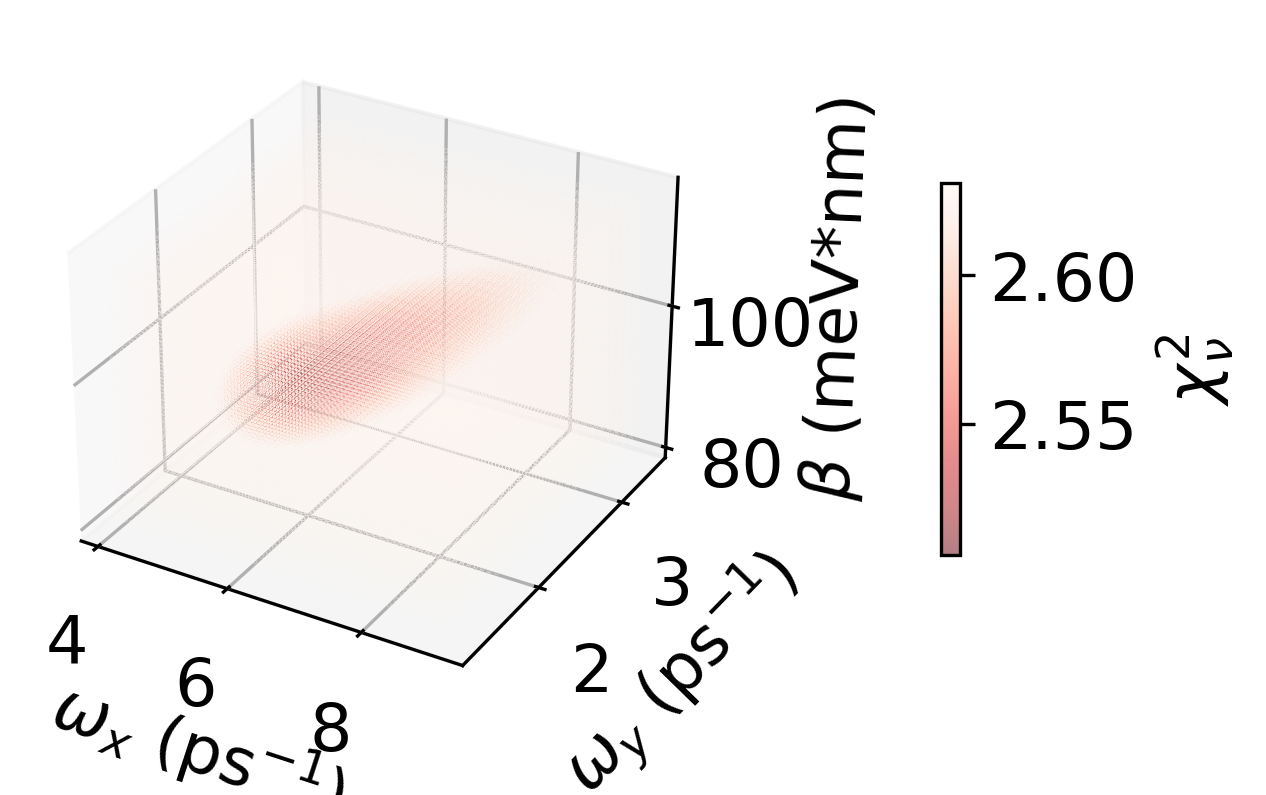}
	\caption{Reduced chi-square $\chi^2_{\nu}$ as a function of the parameters $\omega_x$, $\omega_y$, and $\beta$, showing a region of minimum $\chi^2_{\nu}$. }
	\label{fig:chi_square}
\end{figure}

\subsection{Details of parameters extraction}
\label{sec:extraction}

\begin{figure} [!htbp]
	\centering
	\includegraphics[width=0.72\columnwidth]{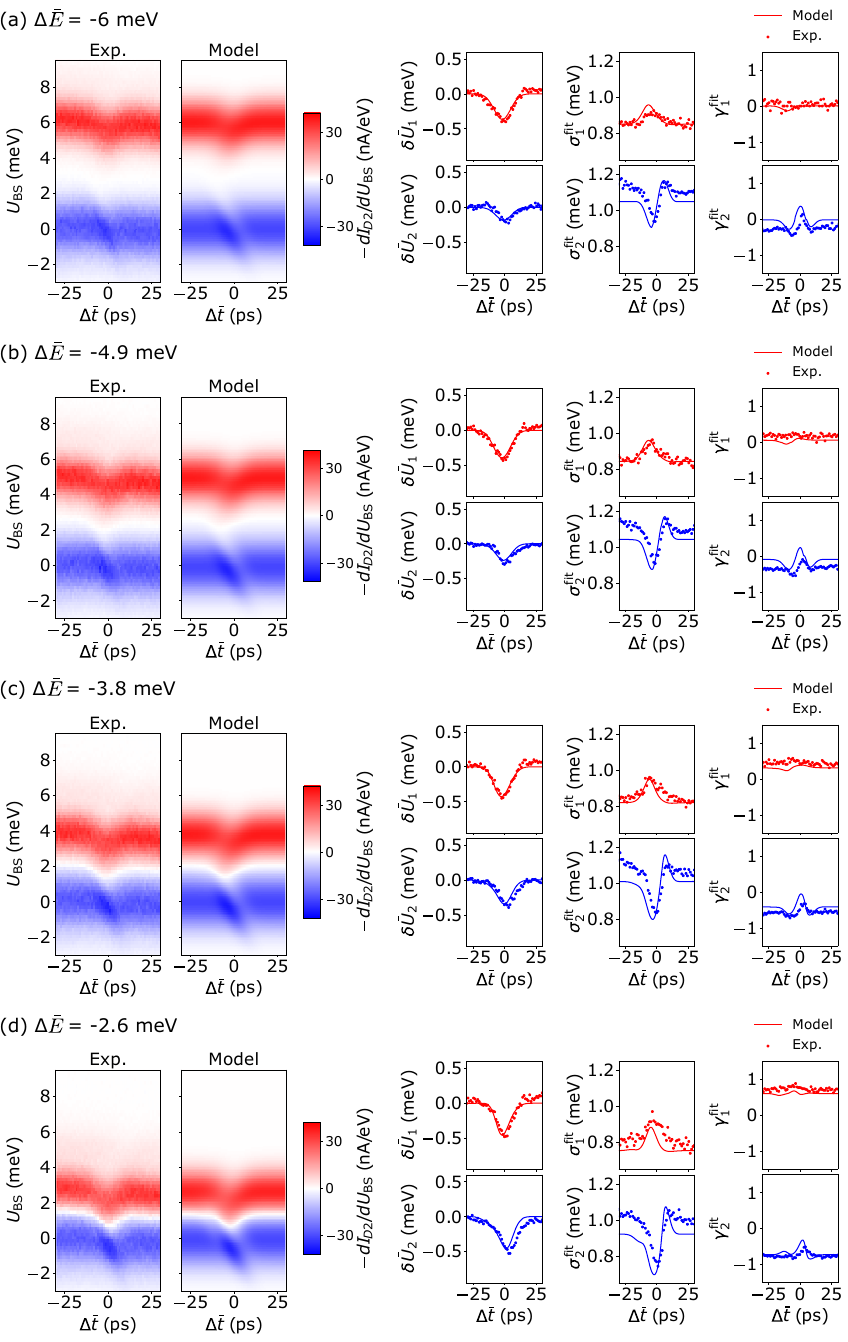}
	\caption{Left panel: Comparison of $dI_{\rm D2}/dU_{\rm BS}$ between the experimental data and the computation result obtained with $\omega_x = 5.8$ ps$^{-1}$, $\omega_y = 2.2$ ps$^{-1}$, and $\beta = 95$ meV$*$nm, at (a-d) $\Delta \bar{E} = -6, - 4.9, -3.8, -2.6$ meV. Right: The difference of the mean value $\delta \bar{U}_i$, standard deviation $\sigma_i^{\rm fit}$, and skewness $\gamma_i^{\rm fit}$. The dots (resp. solid curves) show the values obtained from the experimental (resp. theoretical) $dI_{\rm D2}/dU_{\rm BS}$. The incident energy fluctuation is compensated in computing $\delta \bar{U}_i$ as in Sec.~\ref{sec:fluctuation}. } 
	\label{fig:fitting}
\end{figure}

\begin{figure} [!htbp]
	\centering
	\includegraphics[width=0.72\columnwidth]{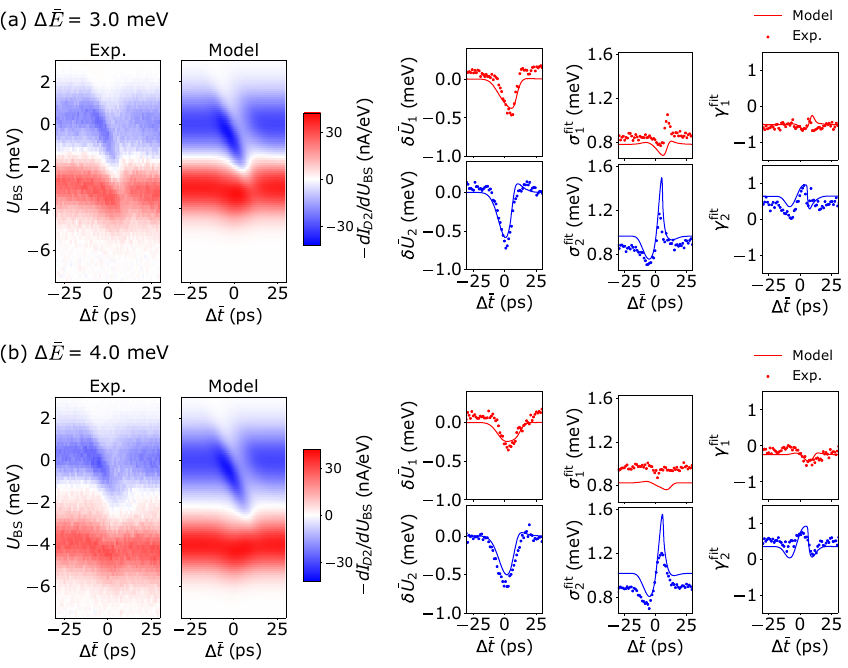}
	\caption{The same as Fig.~\ref{fig:fitting}, but for mean injection energy differences of (a) $\Delta \bar{E} = 3$ meV and $\Delta \bar{E} = 4$ meV.} 
	\label{fig:fitting2}
\end{figure}

We discuss how to extract, from the experimental data, the parameters of the saddle potential curvatures $\omega_x$ and $\omega_y$ and the strength $\beta$ of the Coulomb interaction. We obtain the optimal parameters by finding least square deviation between the experimental $\delta \bar{U}_{i, \rm{ex}}$ and analytical $\delta \bar{U}_{i,\rm{th}}$ [Eqs.~(3)--(4) in the main text], i.e., by finding the minimum chi-square $\chi^2$,
\begin{equation}
\chi^2 = \sum_{i=1}^2 \sum_{j=1}^{N} \dfrac{1}{\sigma_u^2} \big( \delta \bar{U}_{i,\rm{ex}}(\Delta \bar{t}_j) - \delta \bar{U}_{i,\rm{th}}(\Delta \bar{t}_j) \big)^2
\end{equation}
or equivalently finding the minimum reduced chi-square $\chi^2_{\nu} = \chi^2 / \nu$ ($\nu = 2N-3$ equals the number $2N$ of data minus the number 3 of fitted parameters). $\sigma_u$ is the estimated uncertainty in data, which we obtain in Sec.~\ref{sec:error}. To obtain the optimal parameters efficiently, we use the `lmfit' library in python~\cite{Newville2014LMFIT-2}. We find that computed results are insensitive to initial guesses of the parameters, showing that the global minimum of least square presents and is achieved (see also Fig.~\ref{fig:chi_square} for visualization, where $\chi^2_{\nu}$ is computed manually in the parameter space).

\subsection{Comparison between experiment and theory for various values of $\Delta \bar{E}$}

We provide a detailed demonstration that the theoretical model, combined with the extracted parameters ($\omega_x = 5.8$ ps$^{-1}$, $\omega_y = 2.2$ ps$^{-1}$, and $\beta = 95$ meV$*$nm) agrees with the experimental data over various relative mean injection energies of $\Delta \bar{E} = - 6, -4.9, -3.8, -2.6$ meV (Fig.~\ref{fig:fitting}) and $\Delta \bar{E} = 3, 4$ meV (Fig.~\ref{fig:fitting2}). Experimental $dI_{\rm D2}/dU_{\rm BS}$ and the corresponding theoretical values obtained by applying the extracted parameters to the model (described in Sec.~\ref{sec:detector}) are shown in the left panels of Figs.~\ref{fig:fitting} and \ref{fig:fitting2}.

To evaluate agreements between the experimental data and the theory, we compute the mean difference $\delta \bar{U}_i (\Delta \bar{t}) \equiv \bar{U}_i (\Delta \bar{t}) - \bar{U}_i (\Delta \bar{t} \to \infty)$, where $\bar{U}_i(\Delta \Bar{t})$ is the mean value obtained by the integration over finite region of $dI_{\rm D2}/dU_{\rm BS}$. We further compute the standard deviation and skewness by fitting $dI_{\rm D2}/dU_{\rm BS}$ as a function of $U_{\rm BS}$ at each $\Delta \bar{t}$ to the skew-normal distribution function~\cite{azzalini2013skew-2} of $U_{\rm BS}$ with the moments of the mean value $\bar{U}_i^{\rm fit}$, standard deviation $\sigma_i^{\rm fit}$, and skewness $\gamma_i^{\rm fit}$. The skew-normal distribution is chosen to capture the skewness in $dI_{\rm D2}/dU_{\rm BS}$ along $U_{\rm BS}$, caused by the Coulomb interaction. We note that the mean value $\bar{U}_i^{\rm fit}$ obtained by the fitting has almost same behavior with that of $\bar{U}_i$ obtained by the integration.

The results for the mean difference $\delta \bar{U}_i$, standard deviation $\sigma_i^{\rm fit}$, and skewness $\gamma_i^{\rm fit}$ (right panels of Figs.~\ref{fig:fitting} and \ref{fig:fitting2}) show good agreement between the experimental data (dots) and model (solid lines). We note that the experimental $dI_{\rm D2}/dU_{\rm BS}$ slightly deviates from the theoretical $dI_{\rm D2}/dU_{\rm BS}$ for $\Delta \bar{E} = 3,4$ meV (Fig.~\ref{fig:fitting2}) due to the partial emission of the second electron from source S1 with a probability $\sim 20$\%. This happens when we lower the energy of electrons emitted from the quantum dot pump~\cite{Fletcher2013-2}. Overall, our results show the validity of the theoretical model and the extracted parameters over the energy windows of $\Delta \bar{E} \in [-6, -2.6]$ meV and $\Delta \bar{E} \in [3,4]$ meV.

\end{document}